\documentclass[prf,superscriptaddress,twocolumn,sort&compress,round,footinbib,amsmath,amssymb,aps, longbibliography]{revtex4-1}

\usepackage{graphicx}
\usepackage{dcolumn}
\usepackage{bm}
\usepackage{hyperref}
\usepackage[mathlines]{lineno}
\usepackage{float}
\usepackage{mathrsfs}
\usepackage{amsmath}
\usepackage{xcolor}
\usepackage[thinc]{esdiff}

\begin{document}

\preprint{APS/123-QED}

\title{Short-Time Force Response during the Impact of a Droplet with Gas Bubbles}

\author{Edgar Ortega-Roano}
\author{Devaraj van der Meer}
\affiliation{Department of Applied Physics and J.M. Burgers Centre for Fluid Dynamics, University of Twente, P.O. Box 217,
7500 AE Enschede, The Netherlands}

\date{\today}

\begin{abstract}
The presence of gas or vapour bubbles may strongly influence the forces that occur during the impact of a liquid mass onto a solid. Here, we study this effect numerically, in a well-controlled manner, by simulating the short-time interaction between an impacting droplet and a solid surface, mediated by the gas layer between droplet and solid just before collision, in the presence and absence of bubbles. A boundary integral method is used to simulate the falling droplet, the mediating air layer is modeled using lubrication theory, whereas uniform gas bubbles are added to the droplet that obey a polytropic equation of state. We show that the presence of gas bubbles inside the droplet can have a significant influence on the force exerted on the substrate, even before touchdown. This is due to the transmission of load from the solid, through the gas layer and finally into the bubbly droplet, buffering the impact. We simulate different bubble configurations, modifying their number, size, shape and initial position. It is found that larger bubbles, as well as those close to the impact zone, dampen the collision more as compared to small bubbles or the ones that are far from the droplet's surface. In addition, multiple small bubbles are shown to have a similar or even greater effect as a single large bubble.
\end{abstract}

\maketitle

\section{Introduction}\label{sec:2_1}
The impact of a droplet onto a solid surface is an everyday phenomenon experienced many times in our daily life. We have all seen rain drops hitting windows, observed droplets from the shower striking the bathroom walls and floor, or witnessed how water from a leaky faucet falls into the kitchen sink. In many situations, we have felt them impacting our faces and could not help but wonder about the origin of the force that hit us. 

It may come as no surprise that the above processes have been very well studied experimentally, from the shape the droplets take during flight \cite{villermaux2009single}, through their spreading \cite{visser2015dynamics}, possible splashing, rebouncing, and jetting \cite{xu2005drop,xu2007splashing,mou2021singular,zhang2022impact}, to the force they exert on the surface they fall onto \cite{gordillo2018dynamics,mitchell2019transient}. \textcolor{black}{Also 
theoretical progress is extensive, starting with the pioneering work of Wagner about the impact of blunt bodies on a liquid surface \cite{wagner1932stoss} and the pressure profiles these impactors produce onto the liquid, which inspired other seminal articles like the ones from \textcite{lesser1981analytic,korobkin1988initial,scolan2001three,korobkin2006three}, or more recently, the work of \textcite{philippi2016drop}, who provided an expression for the pressure profile for the first moments of the impact following a self-similarity argument.}


Scientists and engineers have also dedicated their time to study erosion and fracturing effects of droplets impacting solids \cite{bowden1961deformation,bowden1964brittle,field2012cavitation,kirols2015effect}, since it is relevant for naval and aviation industry to ensure structural integrity of their vessels 
\textcolor{black}{and} aircrafts. Moreover, special attention has been paid to tackle the droplet impact problem using numerical simulations, predominantly with a volume of fluid (VoF) method \cite{visser2015dynamics,philippi2016drop,zhang2022impact,tretola2022unveiling} or using a boundary integral (BI) approach \cite{davidson2000boundary,hicks2010air,hicks2011air,hicks2013liquid,hendrix2016universal}. This has allowed scientists to access data that is hard to measure experimentally, like the pressure and velocity fields inside and around the droplet. 

Most of the work and examples discussed above encompass the process of droplet impact as a whole. Nevertheless, it is well-known that the largest pressures are occurring precisely at the moment of impact, more hidden from our everyday experience. Even before the touchdown of the droplet, the solid surface feels the impactor's presence due to transmission of momentum from the droplet, through the intermediate gas layer in between and, finally, onto the substrate.  As the droplet approaches the solid, it squeezes air out of its way increasing the pressure in the process. When the pressure in the layer of air is high enough, it is able to deform the droplet's lower surface to create a dimple. This can eventually lead to bubble entrapment at the interface of substrate and droplet \cite{thoroddsen2005air,bouwhuis2012maximal,li2015time,hendrix2016universal}.

Understanding this process is of great importance for the coating, spray and inkjet printing industries, since the entrapment of a bubble will lead to unwanted effects in the printing/coating process \cite{mostaghimi2002splat,mehdi2003air,lohse2022fundamental}. \textcite{hicks2010air} studied air cushioning and bubble entrapment during droplet impact by solving numerically an integro-differential equation describing the droplet's surface and coupling it with a lubrication equation for the air layer beneath the droplet. \textcite{bouwhuis2012maximal} and \textcite{hendrix2016universal} used a BI code to study the maximum air entrapment during the impact of a droplet onto a solid while applying lubrication theory to model the layer of air in between the droplet and solid. Although the focus was on the droplet deformation, the authors were able to compute the pressure inside this layer and, therefore, the load onto the solid surface. They obtained similar pressure profiles to those reported by \textcite{hicks2010air}.

\begin{figure}
\centering
\includegraphics[width=0.43\columnwidth]{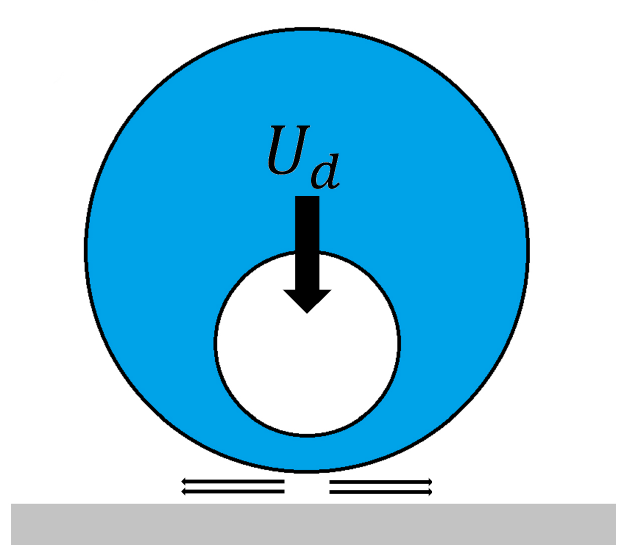}
\makebox{a)}
\includegraphics[width=0.43\columnwidth]{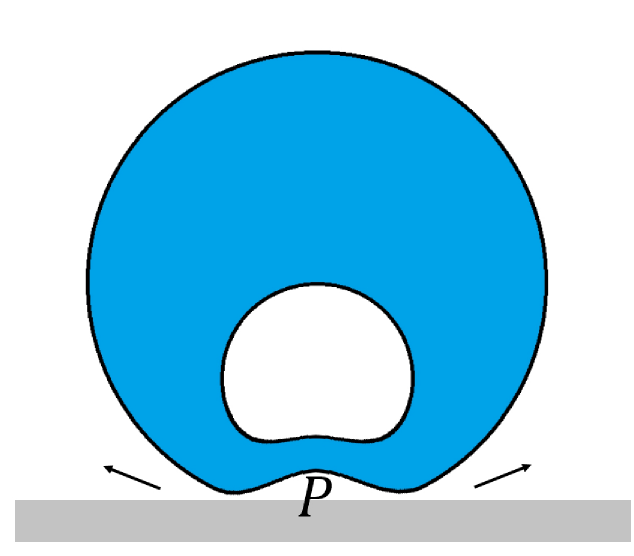}
\makebox{b)}
\caption{ a) An initially spherical droplet containing a bubble falls towards a solid substrate with an initial vertical impact velocity of $U_d$. Beneath the droplet, a layer of gas is squeezed out and escapes from below the droplet while it is falling. b) The pressure inside the layer of air increases until it is large enough to deform the surface of the droplet and bubble. Then a dimple is formed at the bottom of the droplet.
}\label{fig:2_fig1}
\end{figure}

Here, we will show that the presence of gas bubbles inside the droplet has a strong influence on the pressure and forces generated during impact, especially during the short-time response just after impact, where the largest pressures are known to be created, as illustrated in Fig. \ref{fig:2_fig1}. Whereas the longer-time response is typically complex, since the deforming bubble may affect the overall shape of the droplet and possibly include bubble collapse or violent jetting, we will show that already around and just after impact, the presence of a bubble (or multiple bubbles) inside the droplet has a significant effect that does not include large deformations of the bubble but nevertheless strongly influences the load experienced by the solid, typically softening the collision. The position and size of the bubble(s) will be crucial in determining the details of this cushioning effect on the load experienced by the solid, as explained in the following sections.

More specifically, we will study the effect of the presence of one or multiple bubbles inside a droplet just before it impacts a solid substrate. Qualitatively, the process is similar to what happens in the case without a bubble: A droplet with an initial vertical velocity $U_d$ travels down towards a solid surface, Fig. \ref{fig:2_fig1} (a). In between the droplet and the solid, a layer of air is squeezed out with a velocity $u_r$ and with a pressure $P$ that increases as the droplet approaches. Once both solid and liquid surfaces are closer, the pressure in the layer becomes high enough to deform the droplet, indenting its surface. If a bubble inside the droplet is sufficiently close to the surface it will also deform following the deformation of the droplet and the increased pressure in the region of impact, as sketched in Fig. \ref{fig:2_fig1} (b). The pressure in the air and, therefore, the force on the solid will change due to the deformability of the bubble. In the following sections we will explain in detail what differences arise from the case without a bubble and how the position, size and number of bubbles affect the overall short-time response of the droplet during impact.

The paper is structured as follows. In Section~\ref{sec:2_2} we introduce the essentials of the BI method and lubrication theory that will be used in the following sections, as well as the dimensionless parameters that characterize our system. In Section~\ref{sec:2_3} we present the numerical results obtained from the BI. This Section is divided into two subsections, in the first one we deal with droplets that contain a spherical bubble or multiple spherical bubbles only. In the second subsection, we implement bubbles out of the axis of symmetry such that these correspond to toroidal bubbles. In Section~\ref{sec:2_4} we conclude our main findings.

\section{Methods and theory}\label{sec:2_2}
\subsection{Numerical method and lubrication theory }\label{sec:2_2_1}

The Boundary Integral method is used to perform simulations of a bubbly droplet impacting onto a solid. The in-house code used has been tested for different free surface problems in potential flow: a similar code was originally developed by \textcite{oguz1993dynamics} to simulate the detachment of a bubble from a needle, and the in-house code was later adapted by \textcite{bouwhuis2012maximal, hendrix2016universal} to simulate the impact of an inviscid droplet onto a solid, including the air layer between droplet and solid.

\begin{figure}
\centering
\includegraphics[width=0.99\columnwidth]{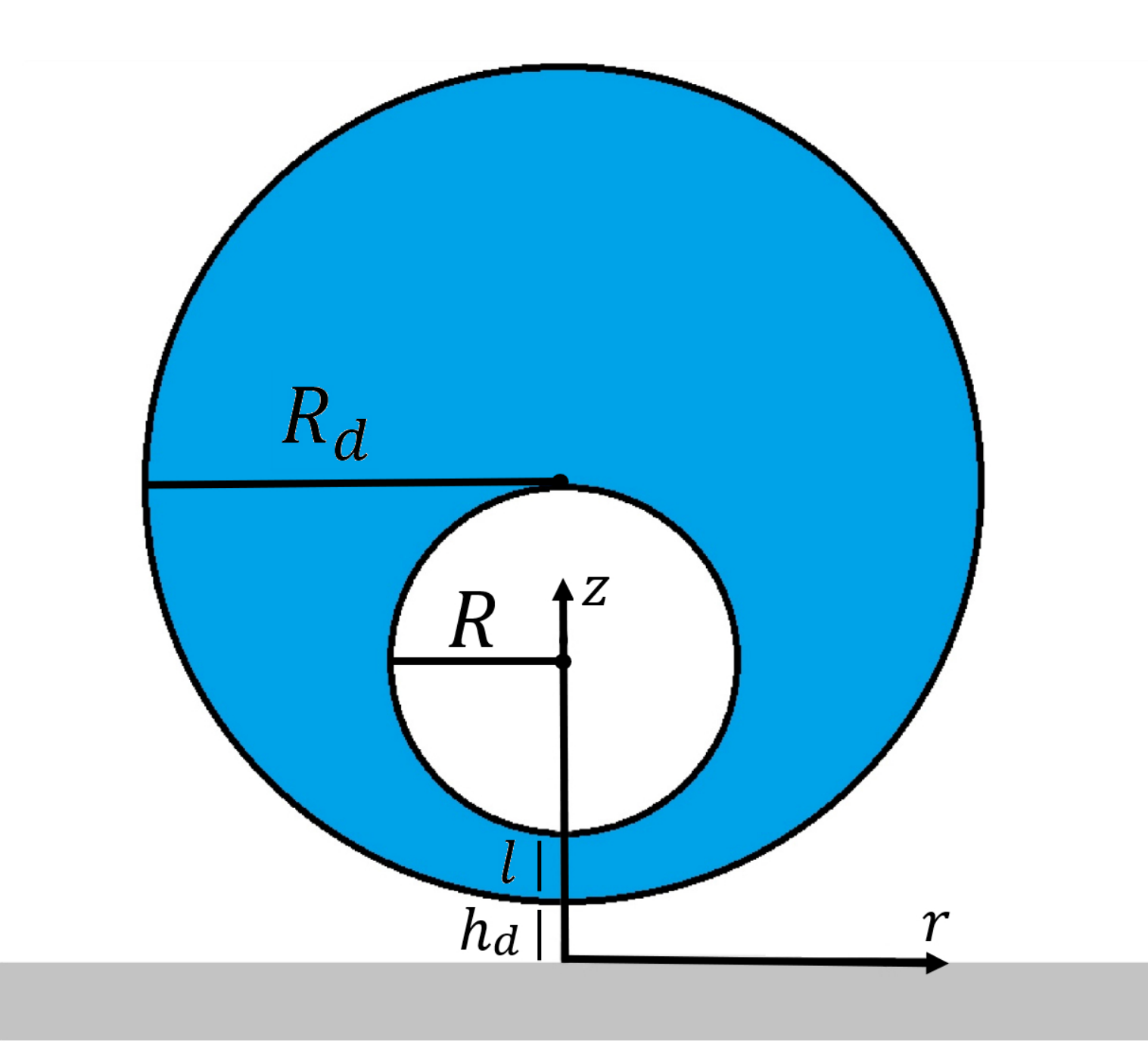}
\caption{ Numerical setup geometry (not to scale): A liquid droplet (in blue) of radius $R_d$ falls towards a solid substrate (in grey). Both the droplet and the very thin air layer between droplet and substrate are included in the simulation, as an inviscid liquid and a viscous lubrication layer for the air, respectively, consistent with earlier work. An air bubble (white) with a radius $R$ is placed inside the droplet, on the vertical symmetry axis and is simulated as a homogeneously pressurized air bubble according to an equation of state, in which flow is neglected. The minimum distance between the bubble and the droplet's surface is given by $l$. The vertical distance between the droplet and the solid substrate is $h_d (r,t)$. The distance $h_d$ never reaches zero since the simulations are stopped just before the droplet impacts on the solid. The cylindrical coordinates are given by $r$ and $z$, the radial and axial coordinate respectively.  
}\label{fig:2_fig2}
\end{figure}

Starting from this code, Fig. \ref{fig:2_fig2} shows a typical geometry of the numerical setup, consisting of a droplet (blue) with one bubble inside (white). Since our BI code is axisymmetrical, $r$ represents the radial coordinate and $z$ the axial coordinate.  The radius of the droplet is given by $R_d$, whereas the radius of the bubble is $R$, $l$ is the minimum initial distance between the bubble and the surface of the droplet. The vertical distance between the droplet's surface and the solid substrate (gray) is labeled as $h_d$. Note that this distance is a function of the radial coordinate and time, i.e., $h_d = h_d(r,t)$, since the droplet will move and deform as the droplet approaches the solid.

The most important assumptions of our liquid solver are that the droplet is inviscid and incompressible. The flow in the liquid is also assumed irrotational such that it is possible to write the velocity field as the gradient of a potential $\mathbf{u} = \nabla \varphi$. It follows that the velocity potential $\varphi$ is an harmonic function, i.e., it satisfies Laplace equation, $\nabla^2 \varphi = 0$. This equation can be solved using Green's method by integrating the potential and its spatial derivatives multiplied by Green's functions over the boundary of the liquid domain. Subsequently the velocity ($\nabla \varphi$) and the time-dependent Bernoulli equation are used to compute the time evolution of the free surface and the flow potential, respectively, after which the Laplace equation is solved again. This is the Boundary Integral method described in detail in Refs. \cite{oguz1993dynamics, gekle2011compressible}.

\textcolor{black}{For the bubble inside the droplet, it is assumed that it is filled with a gas that follows a polytropic equation of state (EOS) \cite{brennen2014cavitation, plesset1977bubble} such that the pressure inside the bubble $p_B$ can be expressed as}
\begin{equation}
    p_B (t) = p_{B,0} \left( \frac{ V_0 }{ V(t) } \right)^{\gamma}
    \label{eq:2_1},
\end{equation}
where $p_{B,0}$ is the initial bubble pressure, $V_0$ its initial volume, $V(t)$ the volume at each time and $\gamma$ is the polytropic exponent. In our simulations $\gamma$ is set to $1$ to model an isothermal bubble expansion and compression, where it should be noted that setting $\gamma$ to other values larger than one will give results similar to the ones presented here. The dynamics of the bubble are coupled to the BI solver via the nodes distributed over the bubble's surface. Clearly, these also contribute to the integration over the boundary carried out in the BI solver. Thus, while the liquid in the droplet is incompressible, the bubble is able to compress and deform following the deformation of the liquid inside the droplet.

A lubrication approach is used in order to model the layer of air in between the falling droplet and the solid, similar to the one employed in \textcite{bouwhuis2012maximal, hendrix2016universal} for the simulation of a droplet impacting onto a solid without bubbles. The reasoning behind using such an approximation is that very close to impact the air layer becomes very thin such that the largest pressures arising are due to the viscous pressure gradient in the layer \cite{davidson2000boundary,hicks2010air,hicks2011air}. Under the lubrication approximation and following \textcite{oron1997long}, it is possible to write the axisymmetric Navier-Stokes equations as 
\begin{equation}
    \diffp{p}{r} \simeq \mu_g 
    \diffp[2]{u_r}{z} \,\,, \label{eq:2_2}
\end{equation}
where $\mu_g$ is the dynamic viscosity of air. After solving \ref{eq:2_2} for the air radial velocity $u_r$, applying the relevant boundary conditions and using continuity, we arrive at an expression that relates the pressure gradient with the vertical position of the droplet $h_d (r,t)$, its rate of change $\dot{h}_d \equiv \partial h / \partial t$ and the surface velocity of the droplet that is in contact with the bottom air layer $u_d$
\begin{equation}
    \diffp{p}{r} = 
    \frac{ 12 \mu_g }{ r h_d^{3} } \int_{0}^{r} \tilde{r} \dot{h}_d \mathrm{d} \tilde{r} + \dfrac{ 6 \mu_g u_d }{ h_d^2 } \,\,, \label{eq:2_3}
\end{equation}
see Appendix \ref{sec:2_5} for a more detailed derivation. Then, it is possible to integrate Eq. \ref{eq:2_3} numerically in the radial direction to obtain the pressure in the air layer and use this expression in the time-dependent Bernoulli equation in the BI solver to evolve the velocity potential in time 
\begin{equation}
    \left( \diffp{\varphi}{t} + \dfrac{1}{2} |\nabla \varphi|^2 \right) = 
     - \dfrac{\sigma}{\rho_l} \kappa - \dfrac{p - p_{\infty}}{\rho_l} ,
\end{equation}
where $\rho_l$ is the droplet's density, $\kappa$ the local surface curvature \cite{bouwhuis2012maximal} and $\sigma$ the surface tension of the liquid-air interface. It is important to note that lubrication theory tells us that $p = p(r,t)$ and does not depend on the axial coordinate $z$, which means that the pressure is constant across the height of the film for any given time and radial distance.  

To obtain the appropriate boundary conditions, the pressure surrounding the droplet, $p$, is divided in two parts: top and bottom. The top pressure is set equal to the atmospheric pressure $p = p_{\infty}$ from the maximum droplet diameter for the entire droplet interface above it. The bottom pressure is obtained from the lubrication equation \ref{eq:2_3} and is applied from the maximum droplet diameter to the entire droplet interface below. Here, it should be noted that the radial pressure gradient becomes very small if $h_d$ becomes sufficiently large such that results hardly depend on at what exact radial position the condition $p = p_{\infty}$ is applied.

\textcolor{black}{Note that we use an incompressible formulation of the lubrication layer in \eqref{eq:2_3}, whereas one may argue that a compressible formulation would be more appropriate, since we are interested in compressibility effects. One should realize however that the pressure build up in the lubrication layer is dominated by viscous effects regardless of the compressibility of the layer. We have also implemented a compressible formulation of the lubrication layer in our code, but found the code to be less stable and the differences with the incompressible formulation to be negligible for the parameter settings used in this work.}

\subsection{Dimensionless parameters}\label{sec:2_2_2}

For definiteness we took our simulated droplet to have a radius $R_d = 0.9 \; \mathrm{mm}$ and to be composed of ethanol, such that its density and surface tension are $\rho_l = 789 \; \mathrm{kg/m^3}$ and $\sigma = 22.27 \; \mathrm{mN/m}$ respectively. Since the BI code is inviscid by construction, the dynamic viscosity of ethanol is of no consequence to the simulations. However, the air layer is considered viscous with a dynamic viscosity $\mu_g = 1.82 \times 10^{-5} \; \mathrm{Pa.s}$. The atmospheric pressure is set to $p_{\infty} = 101.325 \; \mathrm{kPa}$.

Gravity is neglected and thus set to zero. This can be justified by the fact that the Froude number ($\mathrm{Fr}$) for our setup is much higher than one $$\mathrm{Fr^{-1}} = \frac{g R_d}{U_{d}^{2}} \ll 1 .$$
For the simulated droplet initial velocities, ranging from $U_d = 0.2563 - 25.63 \; \mathrm{m/s}$, the inverse Froude becomes $\mathrm{Fr^{-1}} = 1.3\times 10^{-1} - 1.3\times 10^{-5}$, so that using $g = 0 \; \mathrm{m/s^2}$ is a good approximation.

Although surface tension is included in the simulations, inertial effects dominate over surface tension ones. This can be seen if we compute the Weber number ($\mathrm{We}$)
$$\mathrm{We} = \frac{\rho_l R_d U_{d}^{2}}{\sigma},$$
which for the range of velocities given becomes $\mathrm{We} \sim 2 - 2 \times 10^{4}$. In particular, we note that $\mathrm{We} \approx 2 \times 10^{2}$ when $U_d = 2.563 \; \mathrm{m/s}$. This will be the initial droplet speed used for most simulations with the exception of those in Section \ref{sec:2_3_1_2}, Fig. \ref{fig:2_fig7}. The reason for this peculiar velocity choice is that, if we define the Stokes number as 
$$\mathrm{St} = \frac{\rho_l R_d U_d}{\mu_g}, $$
then $\mathrm{log_{10}(St)}$ ranges from 4 to 6, and for $U_d = 2.563 \; \mathrm{m/s}$ it becomes exactly $\mathrm{log_{10}(St) = 5}$. The Stokes number compares the effect of viscous forces inside the air layer and inertial forces from the droplet, which is relevant to describe dimple formation and, as a consequence, to determine the pressure in the air film. For such high Stokes number we are in the so called inertial regime \cite{bouwhuis2012maximal, hendrix2016universal} for air bubble entrapment as opposed to the capillary regime that exists for very low impact speeds.

Note that some authors \cite{bouwhuis2012maximal, hendrix2016universal} relate the capillary number (in the gas phase) to We and St by simply defining $\mathrm{Ca} = \mathrm{We}/\mathrm{St}$. However, this definition is slightly dangerous since one may argue that not $U_d$ but the considerably larger $u_r$ is the typical velocity in the gas phase. To avoid this problem we decided to characterize our simulations using the Weber and Stokes numbers. The Reynolds number of the droplet, defined as $\mathrm{Re}_l = \rho_l R_d U_d/\mu_l$, is assumed to be much larger than one, consistent with the use of the inviscid BI method. As for the Reynolds number of the air layer, $\mathrm{Re}_g = \rho_g h_d u_r/\mu_g$, it is concluded to remain small (order of $10^{-1}$ or smaller) during the entire impact process, \textcolor{black}{ since, by continuity, $u_r h_d \sim U_d \ell $ with $\ell \sim \sqrt{h_dU_d}$ we have $\mathrm{Re_g} \sim \rho_gU_d(R_dh_d)^{1/2}/\mu_g$, which is a decreasing function of $h_d$, i.e., it becomes smaller as $h_d$ becomes smaller in time.}


\textcolor{black}{Finally, we note that the use of dimensionless quantities is useful if the problem is such that the number of free parameters is reduced considerably by non-dimensionalizing. As we will see in the following section, the parameter space for the simulations is huge, and we could only explore a small part of it. Since, naturally, both representations are equivalent, we 
chose to continue the discussion using dimensional quantities.} 

\section{Numerical results}\label{sec:2_3}

\begin{figure}
\centering
\includegraphics[width=0.99\columnwidth]{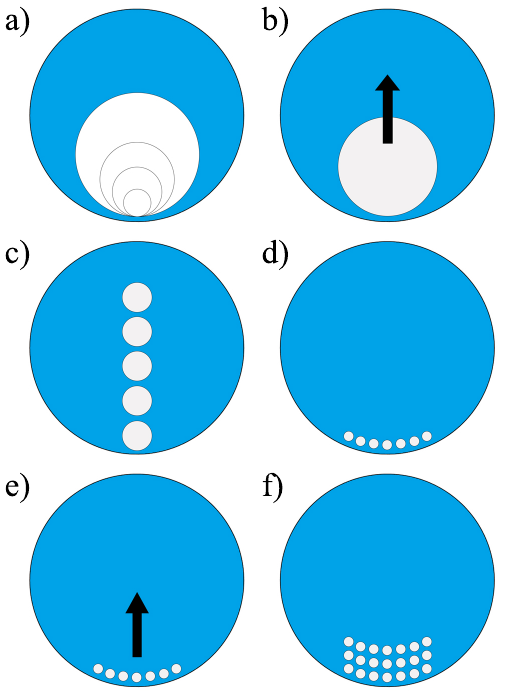}
\caption{Schematic of the different numerical setups that will be discussed: a) Droplet with a single bubble inside with fixed initial distance $l$ and varying radius $R$. b) The same as in (a), but now fixing $R$ and changing $l$. c) Fixing $R$ and $l$, and adding bubbles vertically. d) Fixing $R$ and $l$, adding toroidal bubbles to the sides at a fixed given distance between each other. e) Same as (d), but moving the row of bubbles upwards (i.e., changing $l$). f) Fixing $R$ and $l$, and adding several additional rows of bubbles. Note that bubbles are not necessarily to scale but are representative for each case.}
\label{fig:2_fig3}
\end{figure}

We will now turn to systematically presenting and discussing the results of our numerical simulations. In Fig. \ref{fig:2_fig3} we show the variety of systems simulated, in the same order as they will appear in the following subsections. \textcolor{black}{They can be subdivided into two categories, where the first one only involves spherical bubbles inside the droplet, i.e., bubbles that lie exactly at the axis of symmetry (Sec. \ref{sec:2_3_1}) and the second in addition to spherical ones also includes toroidal bubbles (Sec. \ref{sec:2_3_2}). Since the BI code used is axisymmetrical, any bubble displaced away from the axis of symmetry will automatically become a torus.}

\textcolor{black}{One could argue that the inclusion of toroidal bubbles is dictated by limitations of our numerical model, which is of course true, but arrangements of bubbles entrapped during impact problems are often observed to obey some axial symmetry, even if toroidal bubbles themselves are unstable and would break up quickly by instabilities of the Rayleigh-Plateau or Rayleigh-Taylor type. Such situations have been observed both in experiments \citep{Thoraval2013bubblering} and in 
the modelling of the impact of partially aerated liquids \citep{korobkin2008trapping} or of bodies coated with a thin liquid layer \citep{korobkin2006two}, where the trapping of air between the bodies was discussed.}

Figure \ref{fig:2_fig3} (a) shows the case where we fix the distance between bubble and droplet's surface $l$ and vary the bubble's radius $R$. In (b) we fix $R$ and increase systematically $l$. We then add a vertical bubble array with fixed separation distance between bubbles in (c). 
After that we continue by adding toroidal bubbles to the system, where the center bubble on the axis of symmetry is always spherical. Since the schematics in Fig. \ref{fig:2_fig3} represent cross sections, every torus appears as a pair of circles at the same distance to the left and right from the symmetry axis. In (d) we start adding tori consecutively next to each other with the same distance between them and at the same distance from the droplet's surface. After filling what we call a row of bubbles, we displace it systematically upwards from the surface, as seen in (e). Finally we simulate several rows of bubbles in the vertical direction as shown in (f).

\textcolor{black}{For all the aforementioned cases we use a constant number of nodes to describe the surface of each bubble, namely 
$50$ nodes for spherical bubbles and $100$ for toroidal ones. Since the bubbles can change in volume and deform, the nodes forming them could in theory come very close to each other and eventually overlap, causing the code to stop. In order to avoid this, the nodes are redistributed evenly along natural cubic splines connecting the nodes and describing the bubble surface every second time step.} 

\textcolor{black}{For the outer droplet surface, 
we also redistribute the nodes 
along natural cubic splines every second time step. However in this case, the nodes are not redistributed evenly along the surface but they are allowed to have a minimum and a maximum separation distance, depending on the local droplet curvature. For a droplet of radius 1 (i.e., in dimensionless units), 
the minimum and maximum node distances are $0.001$ and $0.05$ along the surface, 
respectively. This was done in order to obtain more accurate results close to the impact zone, where the deformation of the droplet requires a higher node density (smaller 
node distance), as well as faster simulations since we can simultaneously use less nodes for the upper half of the droplet. Natural cubic splines were used for the redistribution since they smoothly connect the nodes of the free surfaces and they have continuous first and second derivatives.}

\subsection{Spherical bubbles on the vertical symmetry axis}\label{sec:2_3_1}
 
\subsubsection{Fixing $l$, increasing $R$}

This first subsection corresponds to what it is shown schematically in Fig. \ref{fig:2_fig3}(a). We fix the minimum distance $l = 50 \; \mu \mathrm{m}$ of the bubble to the droplet surface (i.e., measured along the vertical symmetry axis) and vary $R$ from 25 to 600 $\mu$m. The droplet is released with an initial $h_d = 10 \; \mathrm{\mu m}$, and a downward vertical velocity $U_d = 2.563 \; \mathrm{m/s}$. This choice of initial $h_d$ is taken such that the excess pressure induced in the air layer by the droplet is negligible as compared to the ambient pressure at the beginning of the motion \cite{bouwhuis2012maximal}.

\begin{figure}
\centering
\includegraphics[trim={25 15 25 15},clip,width=0.99\columnwidth]{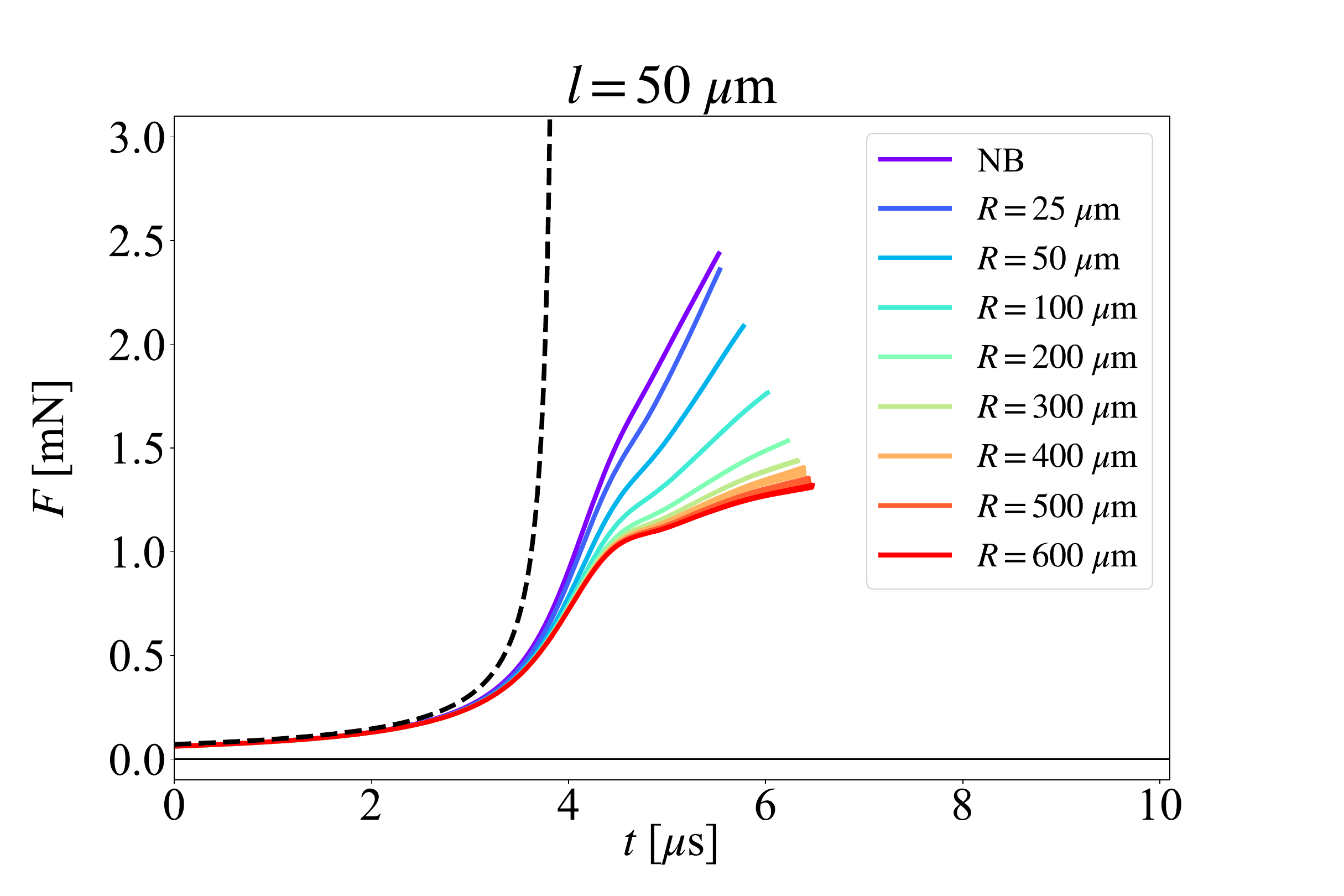}
\makebox{a)}
\includegraphics[trim={15 15 35 15},clip,width=0.99\columnwidth]{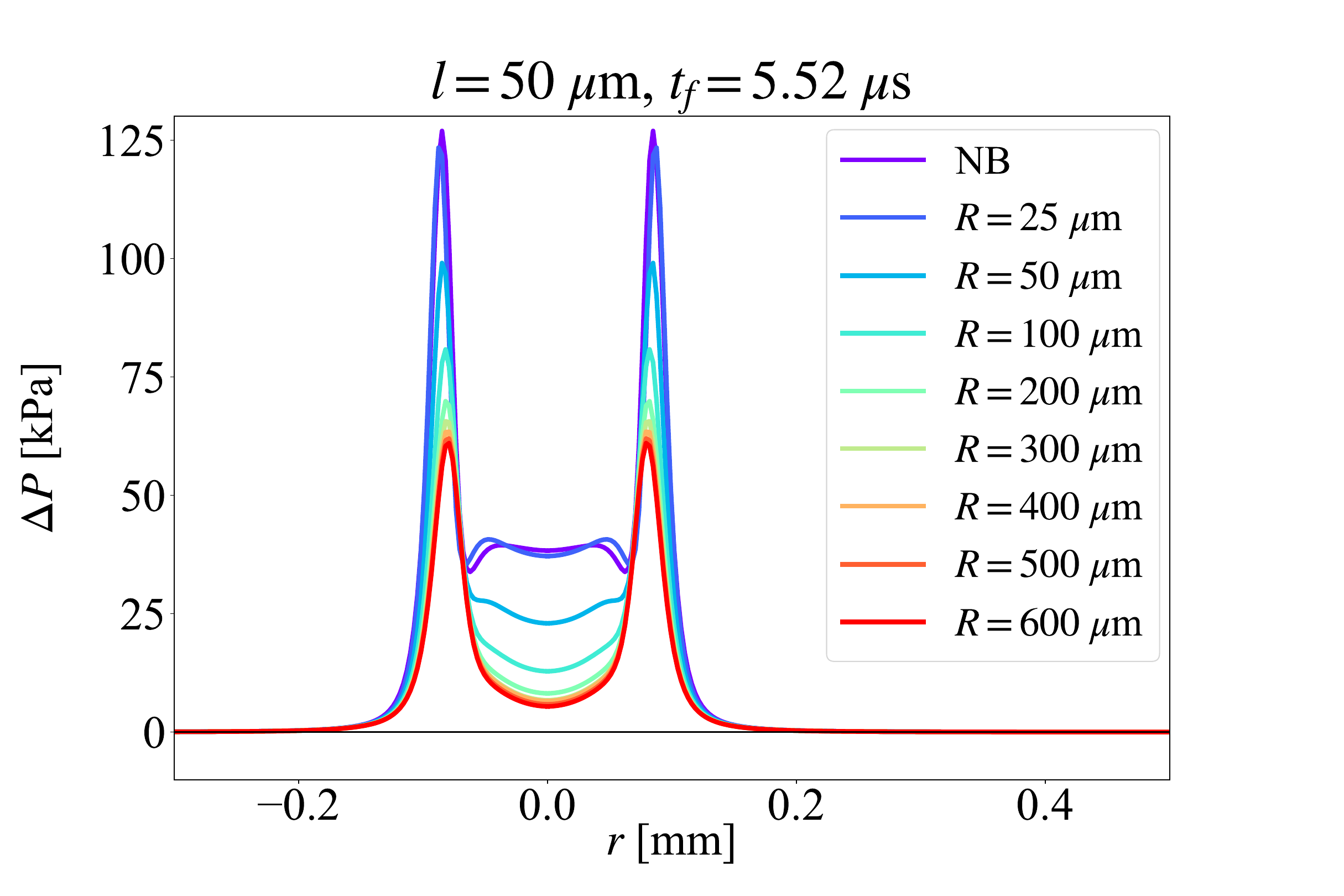}
\makebox{b)} 
\caption{ a) Time evolution of the force $F(t)$ that the solid substrate experiences as the droplet approaches the substrate. The case without a bubble (NB) is represented by the purple line. The other colors correspond to the force for the case of different bubble sizes ranging from $R = 25$ $\mu$m to $R = 600$ $\mu$m. In all the cases the bubble was placed initially at $l = 50$ $\mu$m. The dashed black line corresponds to the theoretical limiting case when the droplet is undeformable (Eq.~\eqref{eq:2_5_2}). b) Pressure profiles $P(r,t_f)$ for the same cases depicted in (a). For comparison, all pressure profiles are plotted at the same time $t_f = 5.52$ $\mu$s, which is the last time step of the NB simulation. }\label{fig:2_fig4}
\end{figure}

Figure \ref{fig:2_fig4}(a) shows the force $F(t)$ exerted onto the solid surface by the impacting droplet as a function of time, where the purple curve represents the case of a purely liquid bubble, i.e., where the droplet has no bubble (NB) inside. As a reference, we will include the NB case as a purple curve in all plots of the following sections. The force was obtained by integrating the pressure in the air layer over the portion of the solid surface lying directly beneath the droplet. Mathematically, this can be expressed as 
\begin{equation}
    F(t) = \int 2 \pi \Delta P(r,t) r \mathrm{d}r \,\,, 
\end{equation}
where $\Delta P(r,t) = p(r,t) - p_{\infty}$ is the excess pressure on the solid surface in the lubrication layer, and $p(r,t)$ is obtained from Eq. \ref{eq:2_3} up until the maximum spreading radius of the droplet. For larger $r$, the pressure on the solid equals the ambient one and, thus, does not contribute to the integral. The simulations are stopped when $h_d = 350$ nm, just before rupture of the air layer and ensuring the validity of lubrication theory given the mean free path in the air layer \cite{bouwhuis2012maximal, hendrix2016universal}. This same criterion is applied in all the following sections.

The different colors in Fig.~\ref{fig:2_fig4}(a) correspond to bubbles of different sizes, with radii ranging from $R=25-600$ $\mu$m, all positioned at a fixed distance $l = 50$ $\mu$m from the droplet surface. There are three main observations to be made by looking at the force: first, for short times all curves overlap, indicating a regime of pressure build up in the air layer that is independent of the presence of the bubble. Second, the larger the bubble is, the smaller the maximum load experienced on the substrate before touchdown and third, larger bubbles delay the touchdown time of the droplet (which correspond to the end point of each curve), i.e., they enlarge what we call the fly time. We will address the first and second points in the subsequent paragraphs and keep the discussion of the third one for next subsection.

Initially, the main response is the kinematic one. The pressure in the air layer has not risen enough to deform the droplet and thus, the added compressibility originating from the bubble plays no role yet. All cases look very similar to the classical squeezed film problem \cite{conway1975impact, bedewi1992squeeze}, where an air film is squeezed from between a plane substrate and an approaching solid sphere. \textcite{bouwhuis2015initial} provide an expression for the pressure inside the air layer for this scenario. It is possible to integrate it in the radial coordinate and arrive at
\begin{equation}
    F(t) = \frac{6 \pi \mu_g U_d R_d^2}{h_{d,0} - U_d t}, \label{eq:2_5_2}
\end{equation}
which is the force experienced by the solid in the limiting case where the droplet can be assumed undeformable. In this expression, $h_{d,0}$ corresponds to the initial droplet height. Equation \ref{eq:2_5_2} is plotted as the black dashed line in Fig. \ref{fig:2_fig4} (a). At around $t = 3 \; \mathrm{\mu s}$ all force curves start to diverge from the undeformable sphere limit and, while the NB simulation already presents some reduction in the load exerted on the solid, the droplets with bubbles inside experience an even greater decrease of the force.

To determine the origin of this decrease, in Fig. \ref{fig:2_fig4} (b) we plot the different pressure profiles $\Delta P (r, t_f)$ over the solid surface for different bubble sizes, all at the same time $t_f = 5.52 \; \mathrm{\mu s}$, which corresponds to the final time of the NB simulation (see the supplemental material \cite{supp} for the pressure profiles plotted instead when the minimum separation distance is 350 nm). First  of all we look at the NB case, for which (as in the other cases) the pressure profile presents some distinct features: The pressure in the center (i.e., on the symmetry axis) is minimal, and the profile presents two maxima which are approximately located at the position where the droplet surface has been deformed. This horn-shaped profile (which of course in fact corresponds to the cross section of an axisymmetric structure) is the hallmark of the Wagner profile corresponding to (inviscid) droplet impact without the interference of an intermediate air layer, as discussed below. For now, we observe that in the presence of a bubble both the central value and the maximum value of the pressure are decreased. This can be understood as follows: The bubble can be compressed and deformed, which makes the droplet interface easier to deform as well. As a result the pressure in the center will rise less compared to the NB case, which in the lubrication layer, also affects the maximum pressure. The effect is stronger as the bubble becomes of comparable size to the footprint of the droplet at the touch down point (with a radius of approximately $\sim 100\,\, \mathrm{\mu m}$), whereas for even larger bubbles the result does not significantly change any further. Clearly, the force is smaller for larger bubbles since the integral of $\Delta P$ is smaller for larger $R$. This can be readily seen by neglecting the axisymmetric three-dimensional character and just qualitatively estimating the area under the $\Delta P$ curves.

\begin{figure}
\centering
\includegraphics[trim={15 15 35 15},clip,width=0.99\columnwidth]{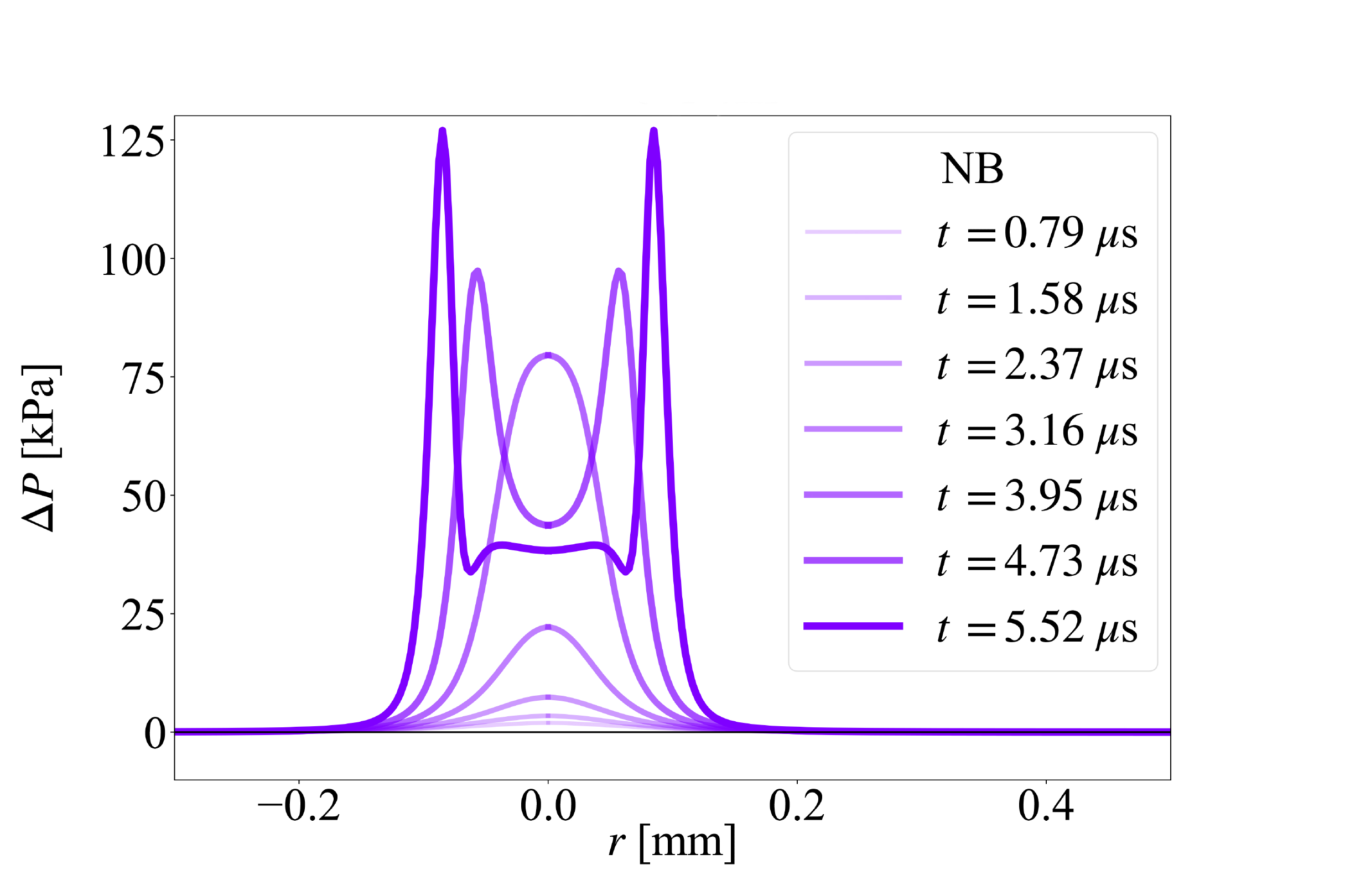}
\makebox{a)}
\includegraphics[trim={15 15 35 15},clip,width=0.99\columnwidth]{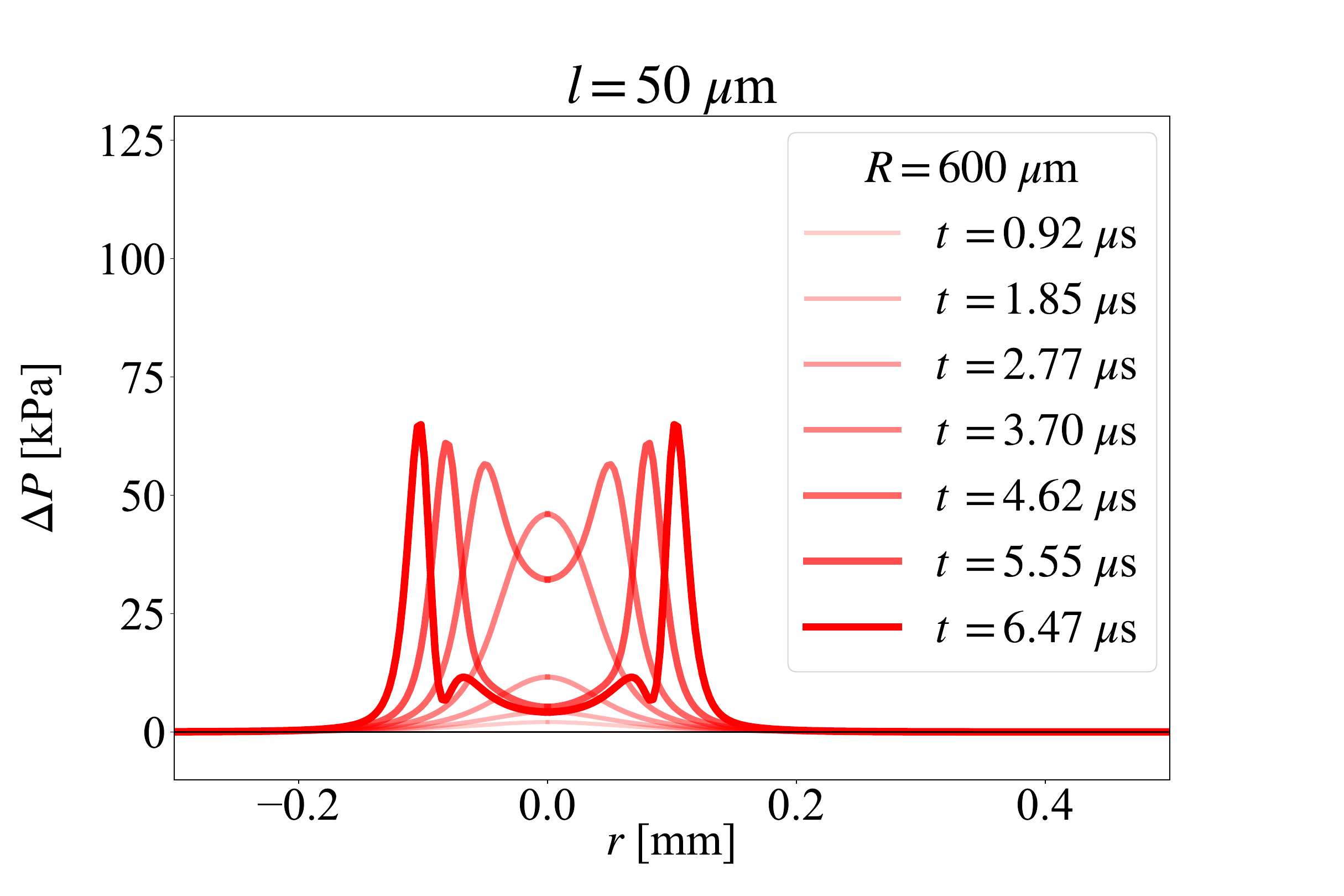}
\makebox{b)}
\caption{ a) Pressure profile $\Delta P(r,t)$ beneath the droplet at different times $t$ when there is no bubble inside of the droplet. The pressure builds up at the center as the droplet falls. Once $\Delta P$ is sufficiently large, the droplet surface will start to deform creating a dimple. The two horns in $\Delta P$ correspond to where the droplet's surface is closest to the solid. b) Pressure profile $\Delta P(r,t)$ at different times $t$ when there is a bubble of radius $R = 600$ $\mu$m inside of the droplet, at an initial distance $l = 50$ $\mu$m from the surface of the droplet. The presence of a bubble decreases the magnitude of the pressure profile while keeping the overall behaviour and shape of the NB case. Note that both axes are the same in (a) and (b).}\label{fig:2_fig5}
\end{figure}

Figure \ref{fig:2_fig5} provides a comparison of the time evolution of $\Delta P$ for the NB and $R = 600 \; \mathrm{\mu m}$ cases. We display seven snapshots of the pressure evenly distributed over their respective maximum fly time. It is seen in both cases that the pressure starts building up from the middle, when the droplet has not deformed yet and the minimum distance between it and the solid lies directly below the droplet's center. The pressure rises, keeping its one-peak shape, corresponding to the squeezed film stage described by Eq.~\eqref{eq:2_5_2}, until the droplet's surface responds and deforms symmetrically creating then the two-horned shape that is also familiar from the purely inertial impact (Wagner profile) \cite{wagner1932stoss,scolan2001three,korobkin2006three}. Once this happens, a dimple in the droplet is formed and the minimum air layer width is no longer at $r = 0$, but starts spreading radially. The minimum transforms then into a ring that delimits the dimple and whose location is exactly at the position of the two peaks present in Fig. \ref{fig:2_fig5} (a) and (b). This is in accordance with what \textcite{hicks2010air,hicks2011air,philippi2016drop,hendrix2016universal} have 
found in their simulations. 

The droplet with a bubble qualitatively shows the same behaviour as the NB one. However, throughout the whole evolution of $\Delta P$, the case with the bubble presents less pronounced peaks. This suggests that the presence of the bubble has a softening effect on the impact. This may well be due to the compressibility the bubble adds to the otherwise incompressible droplet. Not only that, but a bubble close to the droplet's surface is able to deform with ease adapting to the change in shape of the droplet. Part of the cushioning may therefore also be inertial, i.e., a droplet with a bubble inside is simply lighter than a full droplet: the larger the bubble, the less mass is contained inside the droplet.

\subsubsection{Fixing $R$, increasing $l$}\label{sec:2_3_1_2}

We continue with the scenario depicted in Fig. \ref{fig:2_fig3} (b), where we fix the bubble radius to $R = 500 \; \mathrm{\mu m}$ and vary the distance $l$ from the bubble to the droplet surface. This way it is possible to largely exclude the inertial effects discussed above, simply because regardless where the bubble is positioned, the total mass enclosed within the region of interest close to the droplet's surface will be the same. Therefore, any difference in the loads exerted on the solid will most likely be due to the added compressibility and deformability of the bubble.

\begin{figure}
\centering
\includegraphics[trim={25 15 25 15},clip,width=0.99\columnwidth]{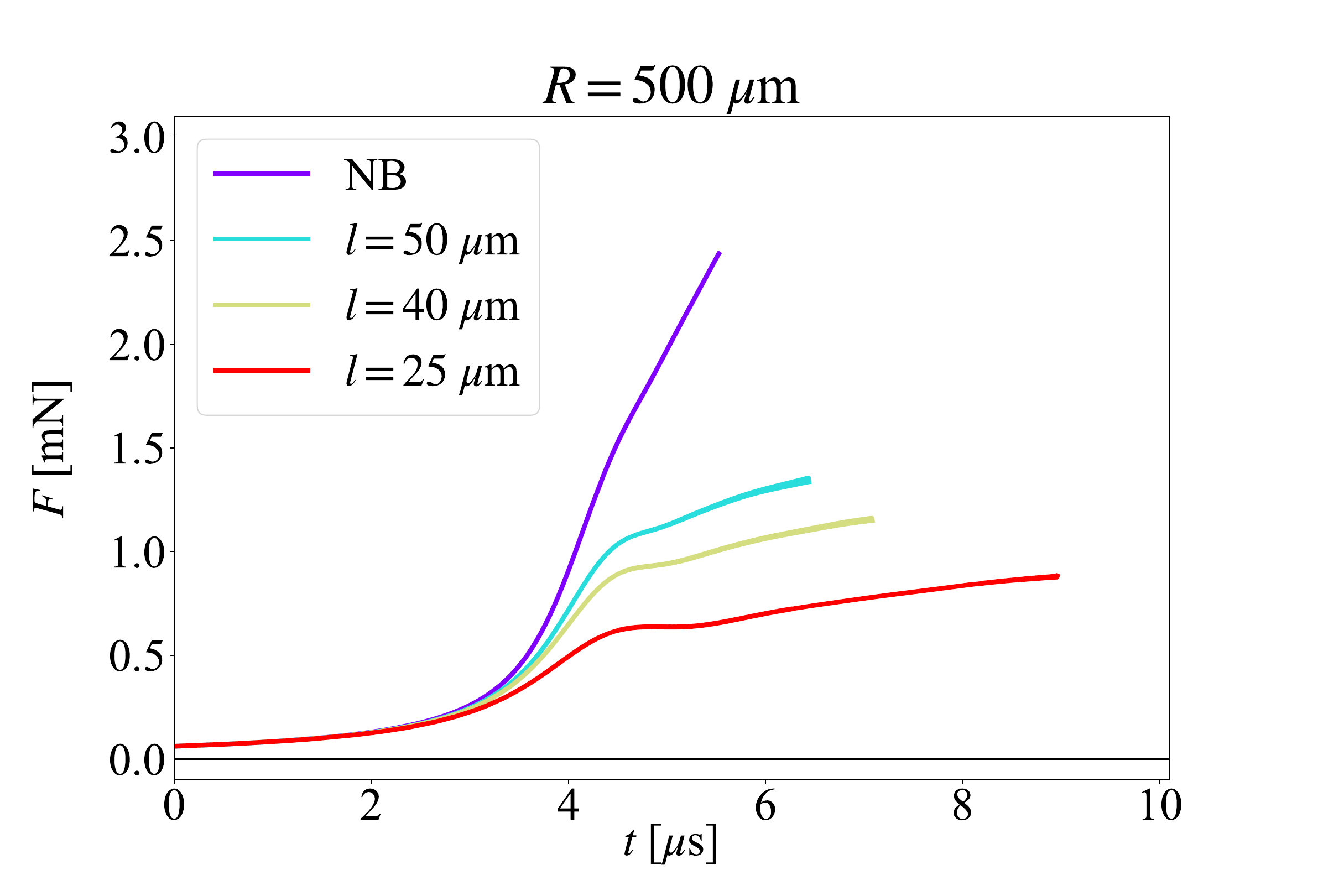}
\makebox{a)}
\includegraphics[trim={15 15 35 15},clip,width=0.99\columnwidth]{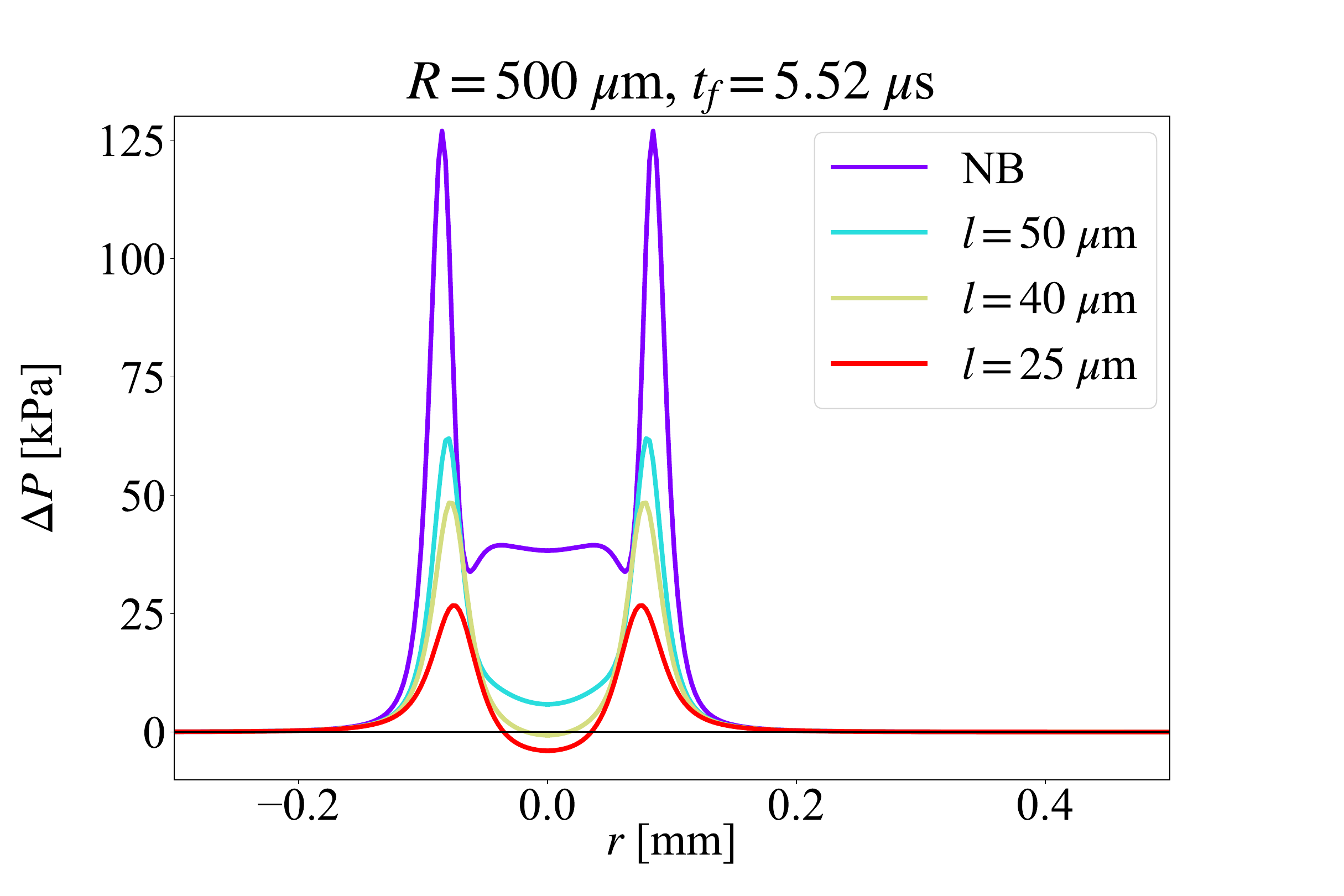}
\makebox{b)}
\caption{ a) Time evolution of the force $F(t)$ on the substrate when fixing the bubble radius to $R = 500$ $\mu$m and varying the distance $l$ of bubble and droplet surface from $50$ to $25$ $\mu$m. Again, the case without bubble (NB) serves for comparison. As $l$ decreases, so does the force exerted on the solid. b) Pressure profiles $\Delta P(r, t_f)$ beneath the droplet for different $l$, plotted at the final time $t_f = 5.52$ $\mu$s of the NB case. Clearly, the smaller $l$ the smaller the pressure becomes. }\label{fig:2_fig6}
\end{figure}

Figure \ref{fig:2_fig6} (a) exhibits the effect of the spherical bubble position on the force. In purple we again plot the NB benchmark case, which presents a higher load for same instances in time and smaller fly time in contrast to the bubbly cases, as previously explained. It is observed that, as the bubble is placed closer to the surface, cushioning is enhanced and the touchdown time increases. Comparing once more the pressure profiles for $t_f = 5.52 \; \mathrm{\mu s}$ (see the supplemental material \cite{supp} for the pressure profiles plotted instead when the minimum separation distance is 350 nm), it is seen that smaller and slightly wider peaks appear for smaller $l$, Fig. \ref{fig:2_fig6} (b). Most significantly, for the smallest distance $l = 25$ $\mu$m, the central pressure even becomes negative, which points to compressible deformation of the bubble.

\begin{figure}
\centering
\includegraphics[trim={5 15 45 15},clip,width=0.99\columnwidth]{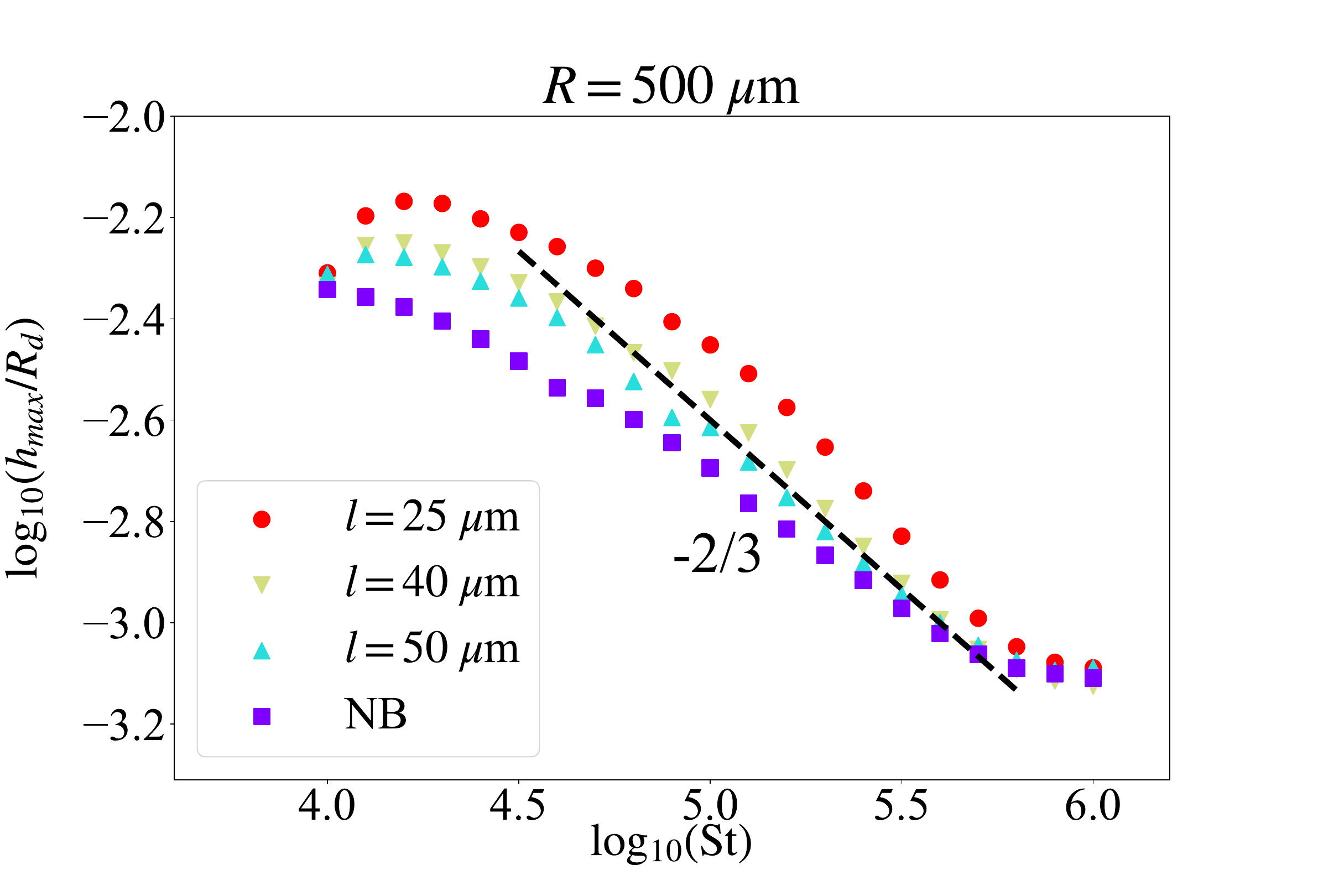}
\caption{ Double logarithmic plot of the normalized maximum dimple height $h_{max}/R_d$ as a function of the Stokes number $\textrm{St}$ in the inertial regime. The purple squares correspond to the case without a bubble and other shapes and colors to a bubble of radius $R=500$ $\mu$m at varying distance $l = 25, 40$ and $50$ $\mu$m from the droplet surface. It can be observed that the dimple height strongly depends on the position of the bubble, and is largest when the bubble is closest to the droplet's surface. For small ($\mathrm{log_{10}(St) = 4}$) and large Stokes number ($\mathrm{log_{10}(St)\sim 6}$) this effect becomes less pronounced. 
}\label{fig:2_fig7}
\end{figure}

In order to quantify the effect of the bubble's position on the maximum dimple height, we perform simulations with initial vertical speeds ranging from $U_d = 0.2563$ to $25.63 \; \mathrm{m/s}$ such that $\mathrm{St} = 10^{4} - 10^{6}$. The simulations are once more stopped when the minimum separation between the droplet's surface and solid is 350 nm. Then, the maximum dimple height $h_{max} = h_d (r = 0, t = t_f)$ is measured for the NB simulation and compared to that of droplets with a $R = 500$ $\mu$m bubble, when $l = 50, 40$ and $25 \; \mathrm{\mu m}$. The results are plotted in Fig. \ref{fig:2_fig7}, where the dimple height normalized with the droplet radius is displayed as a function of the Stokes number. This ranges from $\mathrm{log_{10}(St) = 4}$ to 6, which corresponds to the so called inertial regime of droplet impact. The purple squares correspond to the results for the NB case. For high Stokes number, the data for the NB case should follow \cite{bouwhuis2012maximal}
\begin{equation}
h_{max}/R_d \sim \mathrm{St}^{-2/3}\,\,,  
\end{equation}
and indeed the data tends to converge to this power for large Stokes number. Most importantly however, comparing the case with bubbles to the NB case, we find that the maximum dimple height is larger when the bubble is closer to the droplet surface (smaller $l$) for any Stokes number. This can be explained considering that the presence of a bubble closer to the surface will translate to a higher compressibility in the region influenced by the impact. Since it is possible to deform a bubble with ease, the pressure in the air layer does not need to rise as high as in the NB case in order to deform the droplet. The lower limit $\mathrm{log_{10}(St) = 4}$ indicates the transition to the so-called capillary regime \cite{bouwhuis2012maximal}, where surface tension starts to dominate and tries to preserve the spherical shape of the droplet. All four cases approximately overlap at this point. It is conceivable  that in this regime surface tension prevents the bubble to have a strong effect on the droplet's surface since surface tension starts to dominate over inertia below this Stokes number, which likely is unaffected by the presence of the bubble, especially when realizing that the surface of the bubble interface exhibits the same surface tension as the droplet interface. At the other extreme, for $\mathrm{log_{10}(St) \sim 6}$, the high inertia flattens the bottom of the droplet lowering the dimple height regardless of the presence of a bubble. It is also possible that for high Stokes number a larger temporal resolution is necessary than was used in this work. All simulations were done with a time step of 1 ns, converging in the intermediate Stokes number regime and consequently further research must be done to study to convergence of the maximum dimple height for large Stokes numbers. 

\subsubsection{Fixing $l$ and $R$, increasing number of bubbles}

\begin{figure}
\centering
\includegraphics[trim={25 15 25 15},clip,width=0.99\columnwidth]{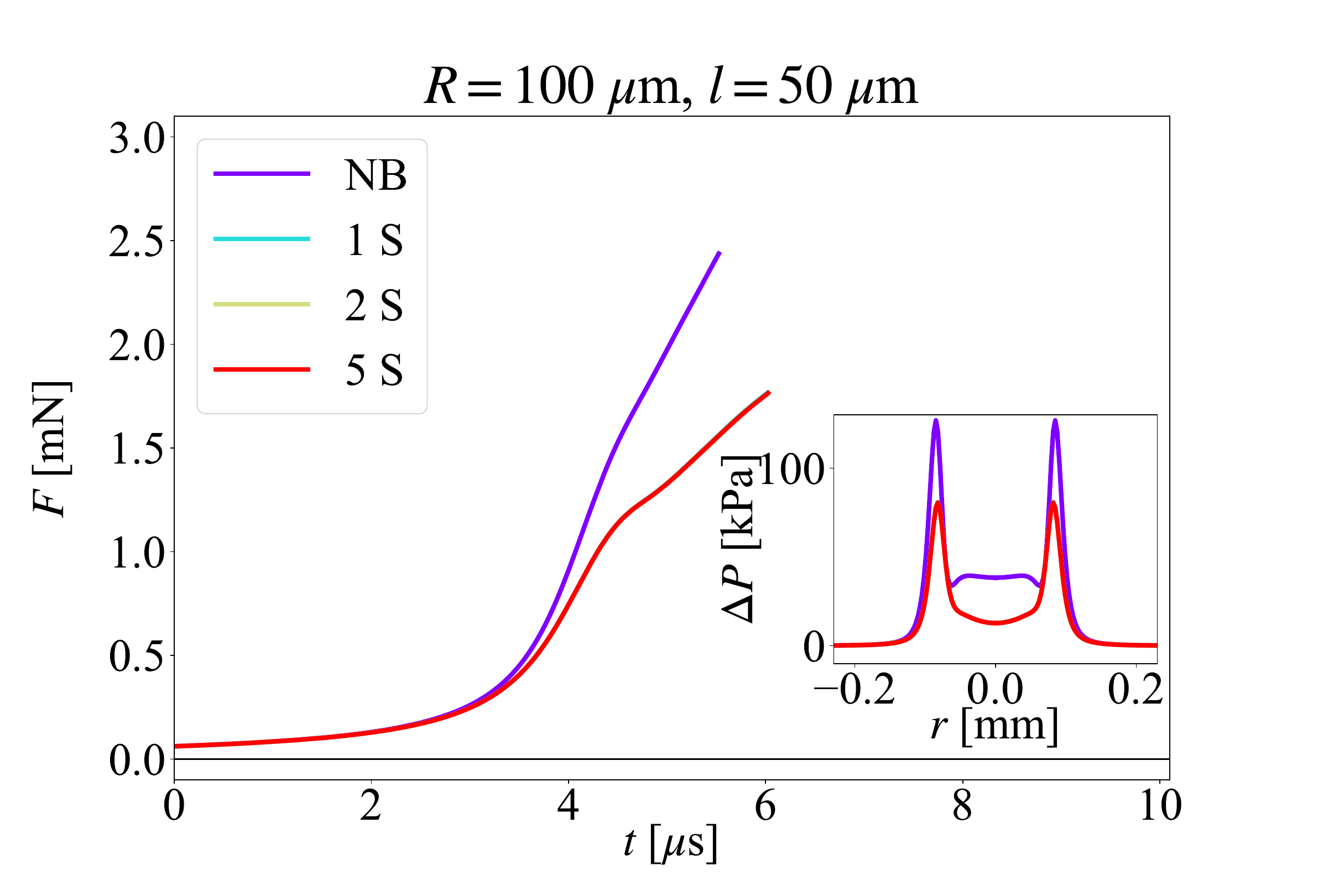}
\caption{Time evolution of the force $F(t)$ on the substrate for different numbers of bubbles stacked vertically as in Fig. \ref{fig:2_fig3} (c). The radii of the bubbles are fixed at $R = 100$ $\mu$m and $l = 50$ $\mu$m. The distance between bubble centers is set to $2R+l$ such that subsequent bubbles surfaces are also at a distance $l$. Again, the case without bubbles (NB) has been added as a reference (purple curve). It is observed that the force response between 1S, 2S and 5S is indistinguishable, and consequently it is only the first bubble closest to the surface that influences the impact. In the inset we show the pressure profiles $\Delta P (r, t_f)$ for different numbers of vertically stacked bubbles, at the final time of the NB case, namely $t_f = 5.52$ $\mu$s. }\label{fig:2_fig8}
\end{figure}

We now turn to the scenario depicted in Fig. \ref{fig:2_fig3} (c). The purpose of this configuration is to investigate the effect of stacking varying numbers of spherical bubbles vertically along the axis of symmetry. Since more than one bubble is simulated, we introduce a new notation. The letter S is used to indicate spherical bubbles and is preceded by a number indicating the number of bubbles, so the case with one spherical bubble is called 1S, with two spherical bubbles 2S and so on. In this case, each bubble has a radius of $R = 100 \; \mathrm{\mu m}$ and the bottom bubble is at a separation distance $l = 50 \; \mathrm{\mu m}$ from the droplet surface. The separation between the centers of the first and second bubble is $2R+l$, such that the minimum initial distance between their surfaces is $l$, and so on up until the last upper bubble (see Fig. \ref{fig:2_fig3} (c)).

The results of these simulations are presented in Fig.~\ref{fig:2_fig8}, from which it is observed that the loads on the solid for the cases 1S, 2S and 5S are practically equal (we do not show 3S and 4S for the same reason). This can also be seen looking at the pressure profiles $\Delta P$ at $t_f = 5.52$ $\mu$m in the inset of Fig. \ref{fig:2_fig8}.
In order to explain this behavior we first take a look at the total liquid volume. The NB case has the largest liquid volume being $3.053 \; \mathrm{mm^3}$, followed by 1S: $3.049 \; \mathrm{mm^3}$, 2S: $3.045 \; \mathrm{mm^3}$ and finally 5S: $3.032 \; \mathrm{mm^3}$. These are comparable to each other, which explains why the bubble cases are identical, but not why they are different from the pure liquid droplet.

One can conclude that the cushioning observed with bubbles inside a droplet is not an effect of the reduced droplet inertia, but a consequence of the positioning of the first bubble, the one closest to the impact zone. Since the first bubble is placed at the same location for all cases (i.e., with 1, 2, 5 bubbles), the reduction in the force is identical.

\subsection{Adding toroidal bubbles}\label{sec:2_3_2}

\subsubsection{Fixing $l$ and $R$, increasing number of bubbles}

In the last subsection we concluded that adding bubbles vertically does not change the behavior. The question that however arises is what happens if additional bubbles are added laterally, i.e., along the bottom surface of the droplet. In order to do so we need to realize that the BI code used is axisymmetric. This means that any bubble displaced away from the axis of symmetry necessarily becomes a torus, which in a cross-sectional image appears as two circles at same height on both sides of and at the same distance from the vertical axis. We now introduce a similar notation as the spherical bubbles where with 1T we denote the case when there is one toroidal bubble, 2T for two toroidal bubbles and so on, usually in combination with spherical bubbles, like 1S 1T or 1S 2T.

\begin{figure}
\centering
\includegraphics[trim={25 15 25 15},clip,width=0.99\columnwidth]{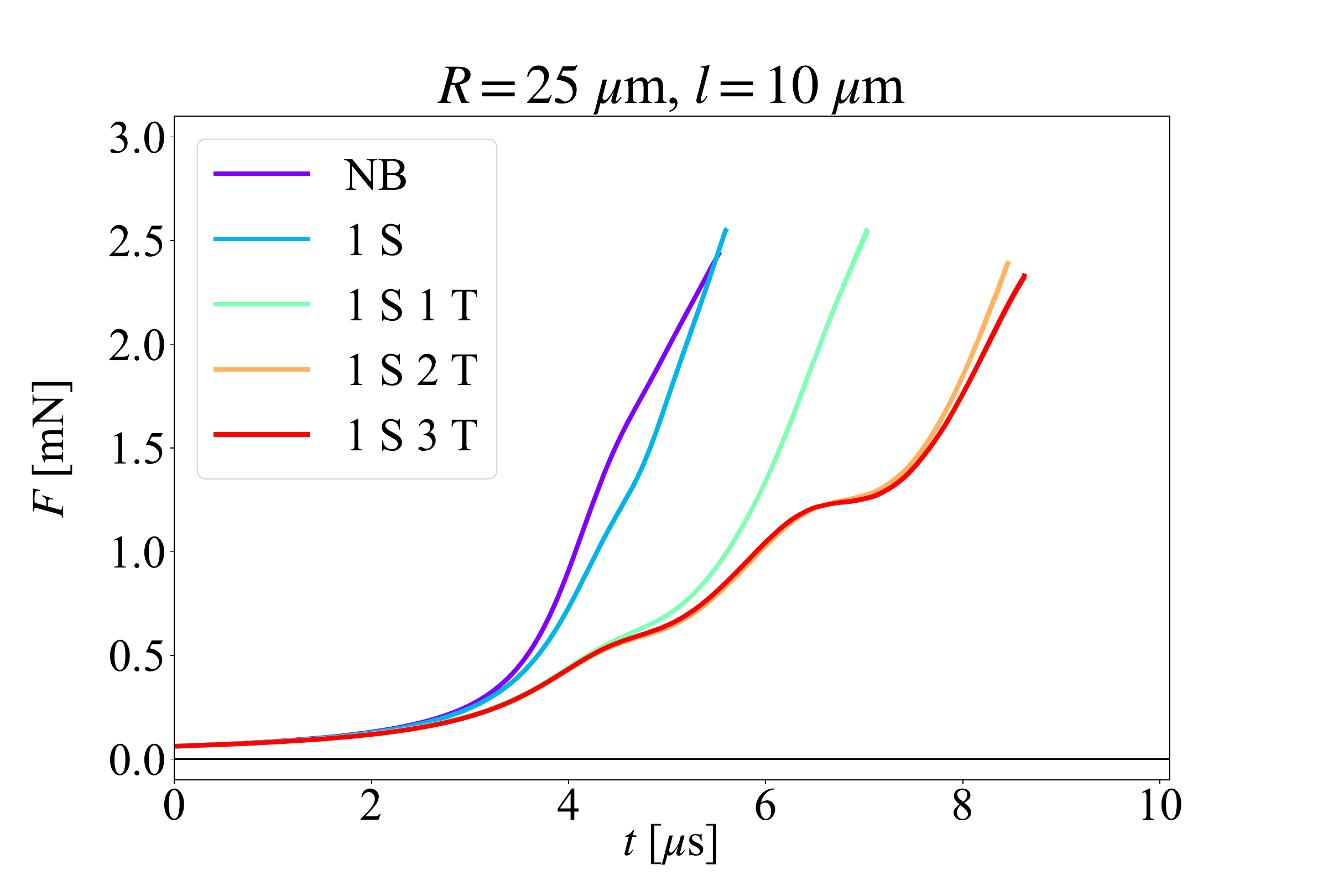}
\makebox{a)}
\includegraphics[trim={15 15 35 15},clip,width=0.99\columnwidth]{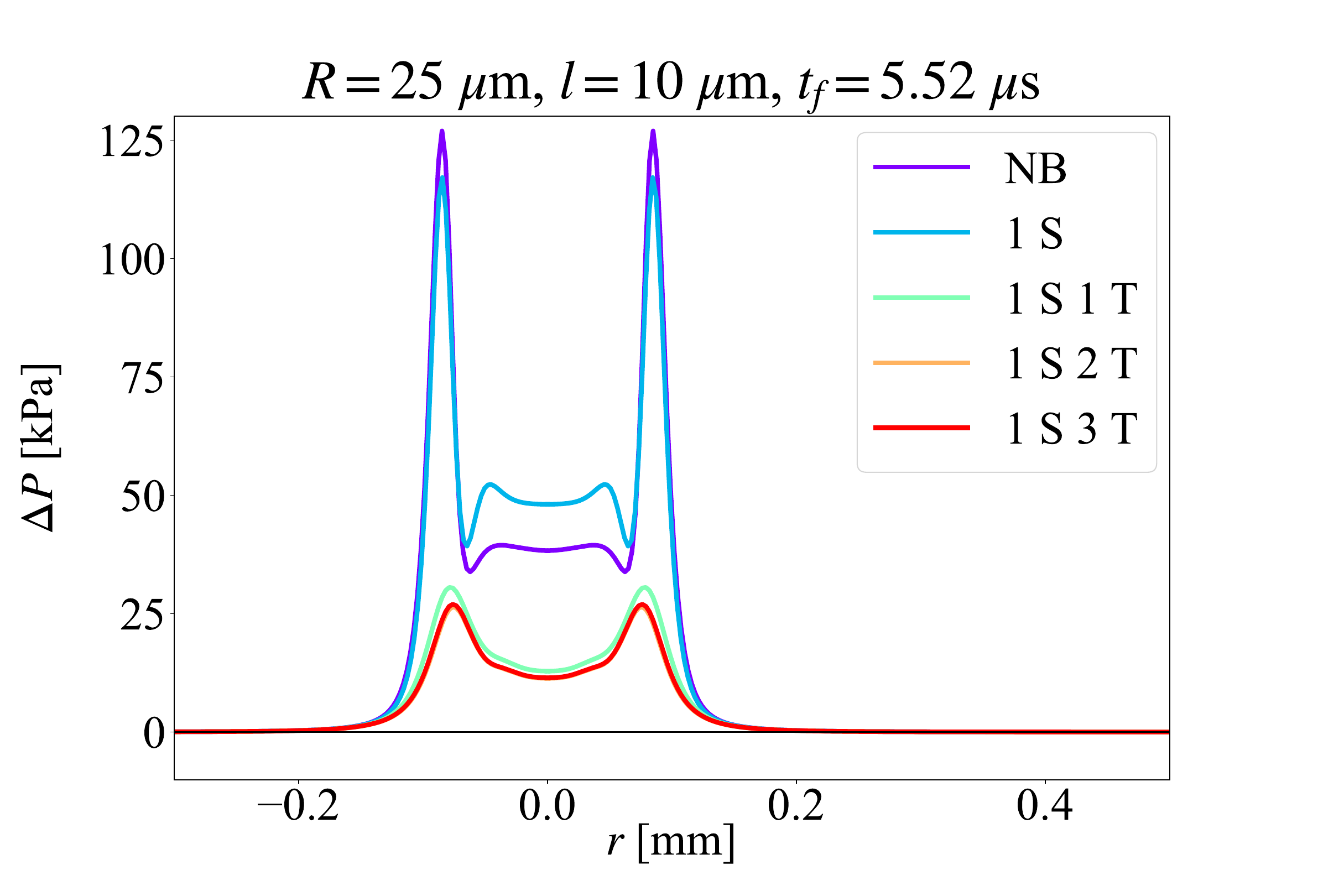}
\makebox{b)}
\caption{ a) Time evolution of the force $F(t)$ exerted on the solid substrate when there are no bubbles (NB), one spherical bubble (1S), and one (1S1T), two (1S2T) and three (1S3T) additional toroidal bubbles (see Fig. \ref{fig:2_fig3} (d) for reference). b) Pressure profile $\Delta P (r, t_f)$ for all these cases, again at $t_f = 5.52$ $\mu$s. It is observed that the presence of more bubbles close to the droplet's surface strongly decreases the pressure on the solid. }\label{fig:2_fig9}
\end{figure}

In this subsection we investigate the scenario of Fig. \ref{fig:2_fig3} (d). The starting case is where a small spherical bubble of $R = 25 \; \mathrm{\mu m}$ is placed close to the droplet's surface at $l = 10 \; \mathrm{\mu m}$, from which we start adding toroidal bubbles to its side for the following simulations. The first torus has a minor radius of $R = 25 \; \mathrm{\mu m}$ (i.e., the same cross section as the spherical bubble), and the minimum distance between its surface and the droplet is $l = 10 \; \mathrm{\mu m}$ as well. We apply an initial separation angle of 0.07 rad between bubbles 
\textcolor{black}{(which is the angle between lines joining the centers of the two bubbles to the center of the droplet)}) in order to place them far enough from one another such that bubbles do not overlap during impact. All other consecutive toroidal bubbles have the same minor radius $R$, \textcolor{black} {the same minimum distance $l$ 
to the droplet's surface 
and are again separated by an angle of 0.07 rad from each other, i.e., the angle between lines joining the centers of two adjacent bubbles to the center of the droplet is 0.07 rad.}

\begin{figure}
\centering
\includegraphics[trim={25 15 25 15},clip,width=0.99\columnwidth]{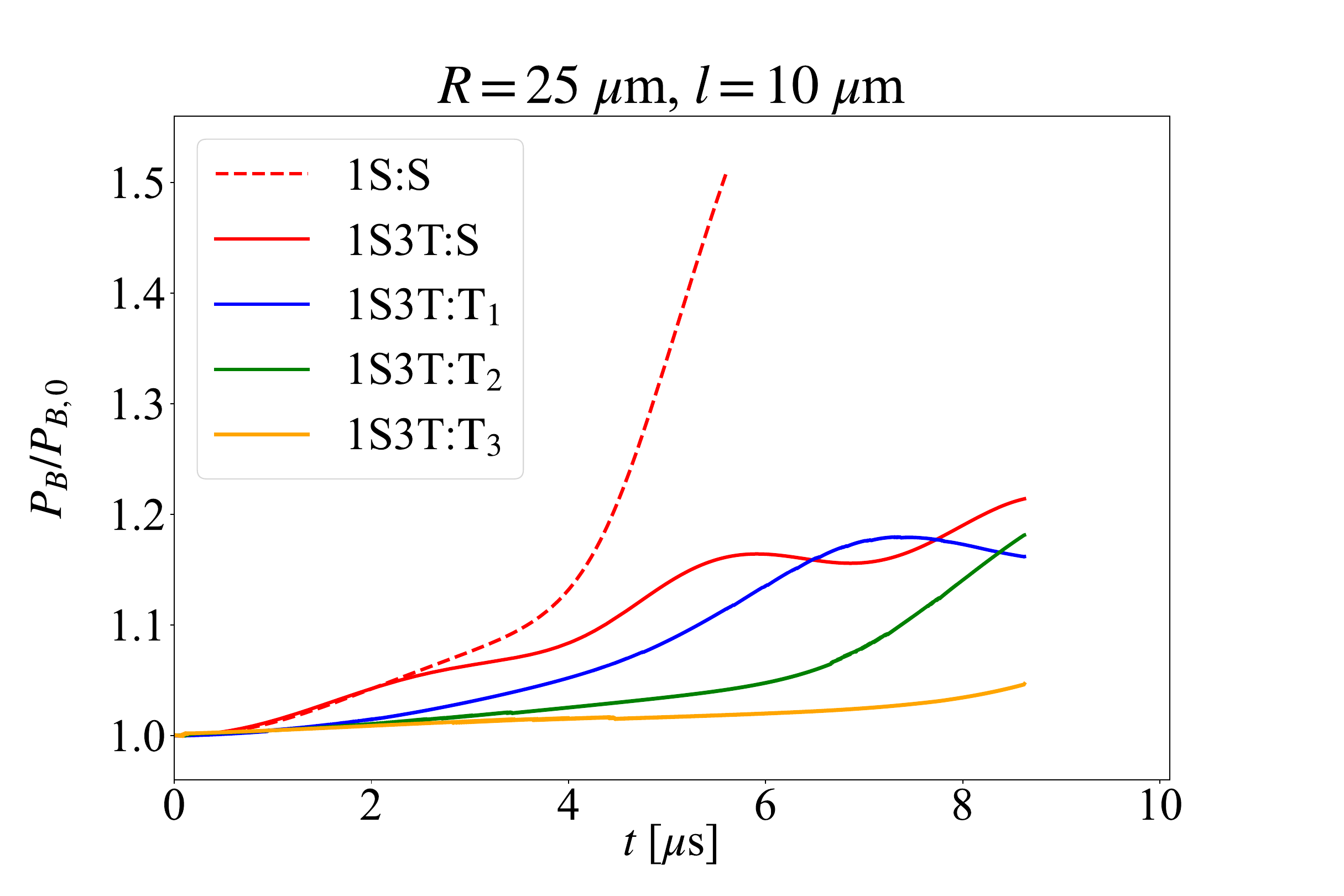}
\makebox{a)}
\includegraphics[trim={15 15 35 15},clip,width=0.99\columnwidth]{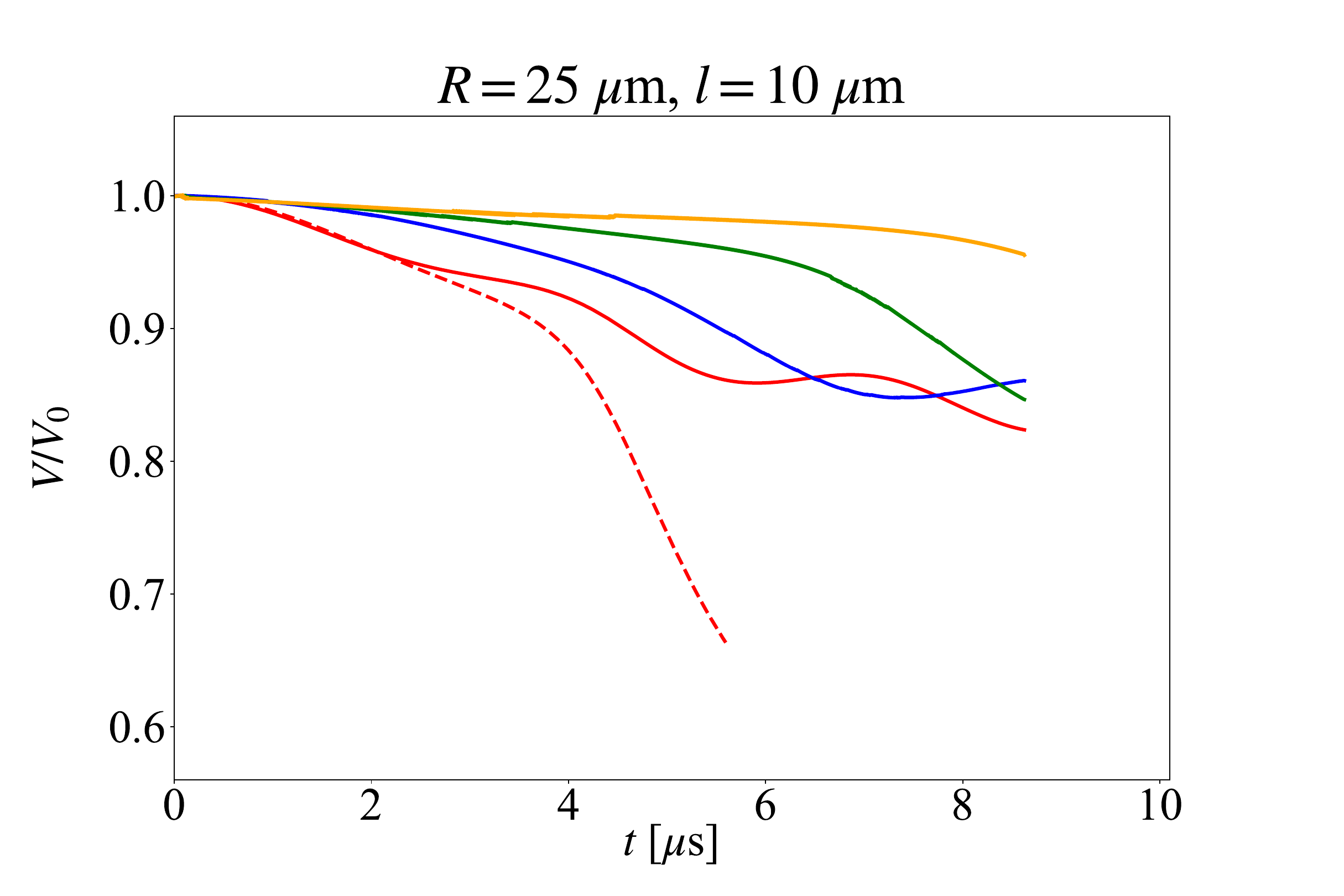}
\makebox{b)}
\caption{a) Pressure inside the bubbles, normalized by their initial pressure, for the simulations of a droplet with a single spherical bubble (1S, dashed line) and for each bubble in a system with one spherical and three toroidal bubbles (1S3T, solid lines). The red curve corresponds to the spherical bubble (S) and the blue, green and yellow ones to the first ($\mathrm{T_1}$), second ($\mathrm{T_2}$) and third ($\mathrm{T_3}$) toroidal bubble respectively, counting from the symmetry axis outwards. b) Volume of the bubbles normalized by their respective initial volume. }\label{fig:2_fig9_1}
\end{figure}

\textcolor{black}{Note that a toroidal bubble with a minor radius $R$ has a substantially larger volume as compared to the center spherical bubble with a radius of the same magnitude. One could choose to run simulations with toroidal bubbles that have the same equivalent volume as the spherical one. However, we decided to take the minor radius of the tori equal to the radius of the spherical bubble in order to have comparable pressures. It should be noted that in our case, 
the compressibility in both the spherical and the toroidal bubbles is 
determined by an air layer of approximately the same thickness.}

Figure \ref{fig:2_fig9} (a) and (b) show the force and pressure profiles for the cases NB, 1S, 1S1T, 1S2T and 1S3T. There are several observations to be done. The first one concerns the differences between NB and 1S. It is seen that in the case without bubble (NB) the droplet exerts an overall larger force on the solid, but the case with one spherical bubble (1S) is not too far off, even overlapping and slightly surpassing the NB case at around 5 $\mu$s. This can be explained if we look at Fig. \ref{fig:2_fig9_1} (a) and (b) where the internal pressure and volume of the bubbles are displayed respectively. The dotted red line corresponds to the simulation 1S. It is seen that initially the bubble undergoes a slow compression (decrease in volume), which subsequently speeds up at around $t = 4 \; \mathrm{\mu s}$, increasing the pressure rapidly in the process. A higher pressure inside the bubble indicates a high resistance to additional compression and deformation of the droplet's surface, therefore increasing the pressure in the air layer, as observed in Fig. \ref{fig:2_fig9}(b) for the 1S case.

The solid lines in Fig. \ref{fig:2_fig9_1} correspond to the simulation 1S3T, containing one spherical and three toroidal bubbles. The colors red, blue, green and yellow represent the spherical bubble (S), first ($\mathrm{T_1}$), second ($\mathrm{T_2}$) and third ($\mathrm{T_3}$) toroidal bubbles respectively. It is evident that the compression of the spherical bubble is less pronounced when it is accompanied by toroidal bubbles. As the droplet falls and deforms, the impact area of influence grows laterally reaching the tori and the deformation is now spread over more bubbles. Looking from a mass conservation perspective, the more bubbles there are in the system, the more places where liquid is able to flow towards with ease when it is being pushed up by the air layer.

Looking back at Fig. \ref{fig:2_fig9} (a), it is observed that the cases with toroidal bubbles present a stronger force attenuation, even compared to 1S. In addition, we notice that the simulations 1S2T and 1S3T are almost identical up until the last microsecond. This is because the impact area of influence reaches the third torus just before touchdown, which is explained in more detail in the next section.

\subsubsection{Fixing $R$ and number of bubbles, increasing $l$}

For brevity, a system composed of one spherical bubble and three consecutive tori (1S3T) will be called a full row. From the results from the last section, it is clear that adding another torus (i.e., creating 1S4T) does not significantly reduce the load on the solid any further, at least in the time frame up to the first contact with the solid that we are focusing on in this Chapter. Therefore, this section is dedicated to explore the effect of displacing a full row (1S3T) further away from the droplet's surface. This is depicted schematically in Fig. \ref{fig:2_fig3} (e). In doing so, the radii of the bubbles are kept the same, $R = 25 \; \mathrm{\mu m}$, again with a constant separation angle of 0.07 rad.

\begin{figure}
\centering
\includegraphics[trim={25 15 25 15},clip,width=0.99\columnwidth]{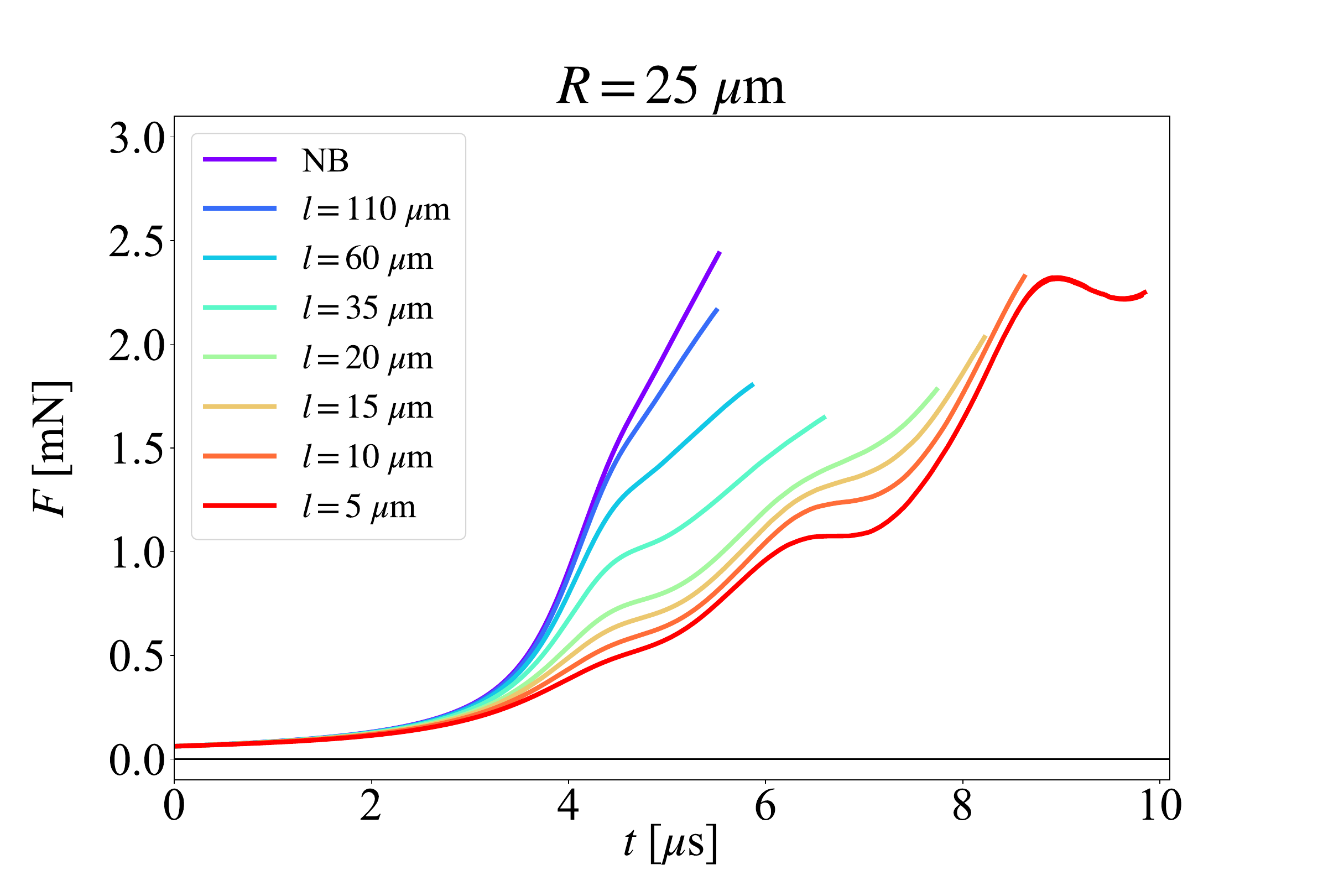}
\makebox{a)}
\includegraphics[trim={15 15 35 15},clip,width=0.99\columnwidth]{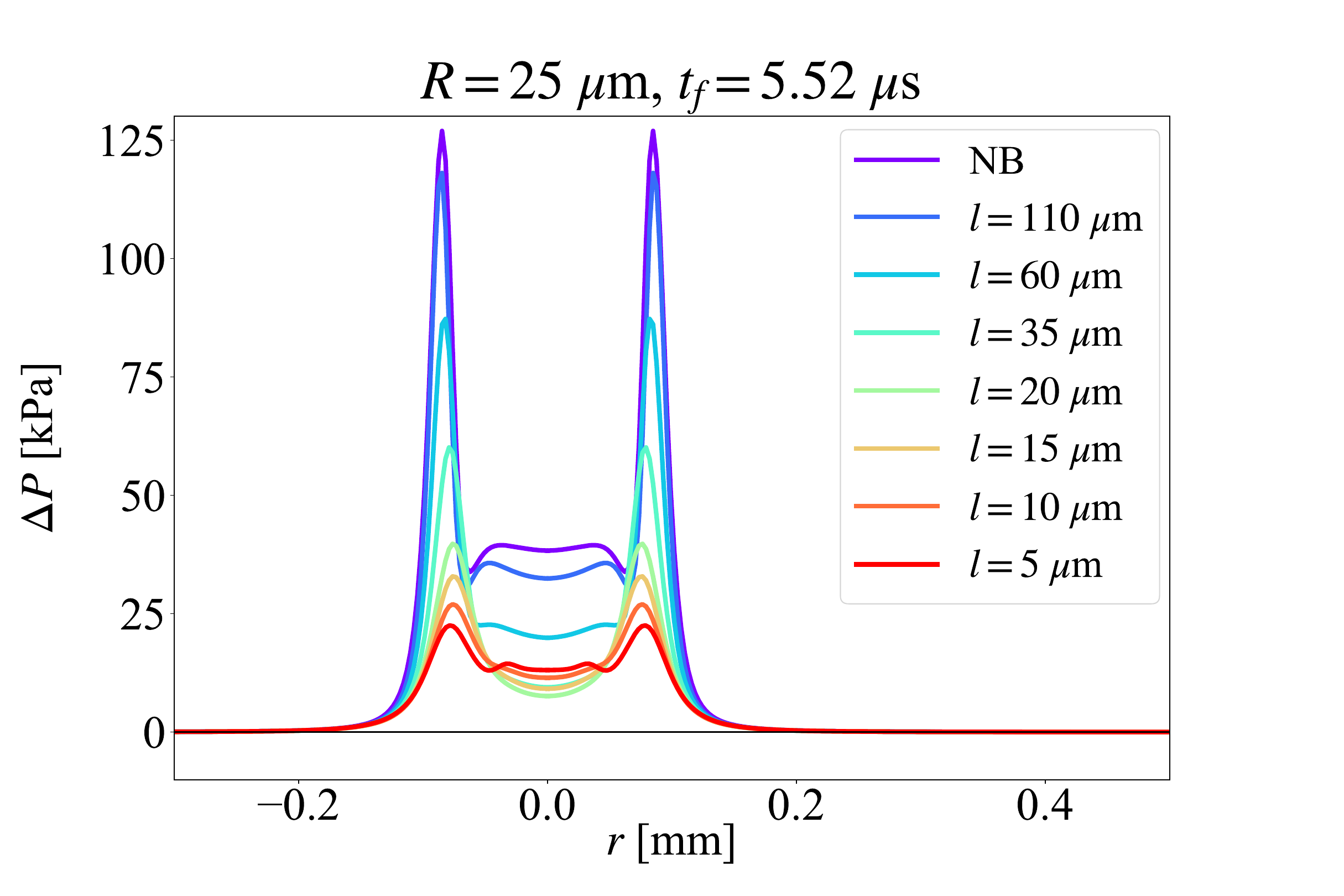}
\makebox{b)}
\caption{ a) Time evolution of the force $F(t)$ exerted on the solid in the presence of a full row of bubbles with $l$ varying from 5 $\mu$m to 110 $\mu$m. The closer the row of bubbles is to the droplet surface, the smaller the force experienced by the solid as compared to the case without a bubble at similar times. b) Pressure profiles $\Delta P (r,t)$ at $t_f = 5.52$ $\mu$s, corresponding to the force curves in (a). }\label{fig:2_fig10}
\end{figure}

Figure \ref{fig:2_fig10} (a) and (b) show once more the force as a function of time and pressure profile at $t_f$ respectively. Again, the curves are color coded, where dark blue represents the case in which the force resembles the NB case the most, corresponding to the largest distance $l = 110 \; \mathrm{\mu m}$. Here, the bubbles are so far from the impact zone that the cushioning from one full row is minimal.

The distance $l$ is reduced with each consecutive simulation until $l = 5 \; \mathrm{\mu m}$ is reached. This particular run presents the strongest force attenuation from this set and the longest fly time from all simulations discussed in this study. An attempt to reduce $l$ even further was made, but the simulations turned out to be unstable due to close proximity of the droplet surface and the sudden increase in pressure in the air layer.

\begin{figure}
\centering
\includegraphics[trim={25 15 25 15},clip,width=0.99\columnwidth]{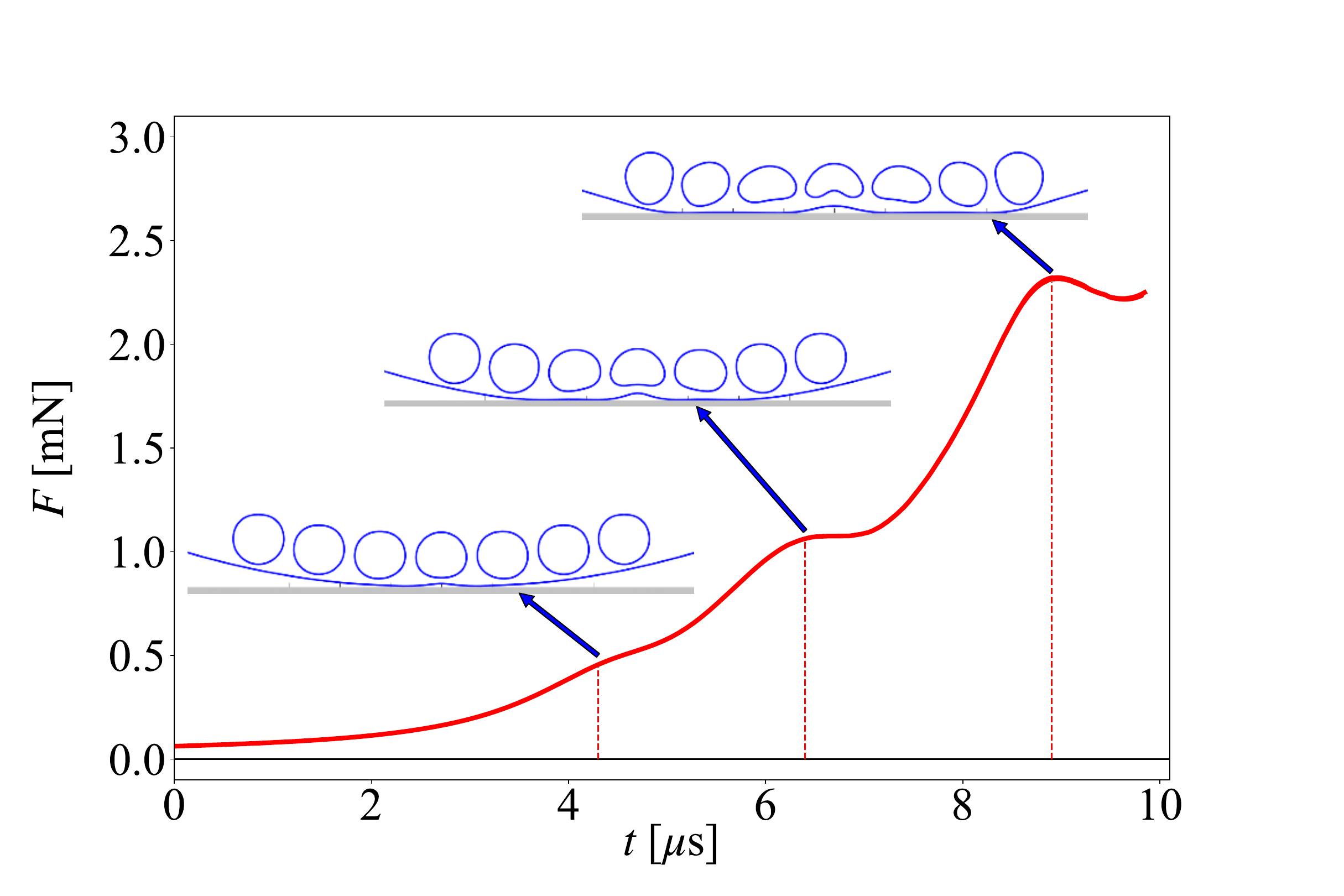}
\makebox{a)}
\includegraphics[trim={15 15 35 15},clip,width=0.99\columnwidth]{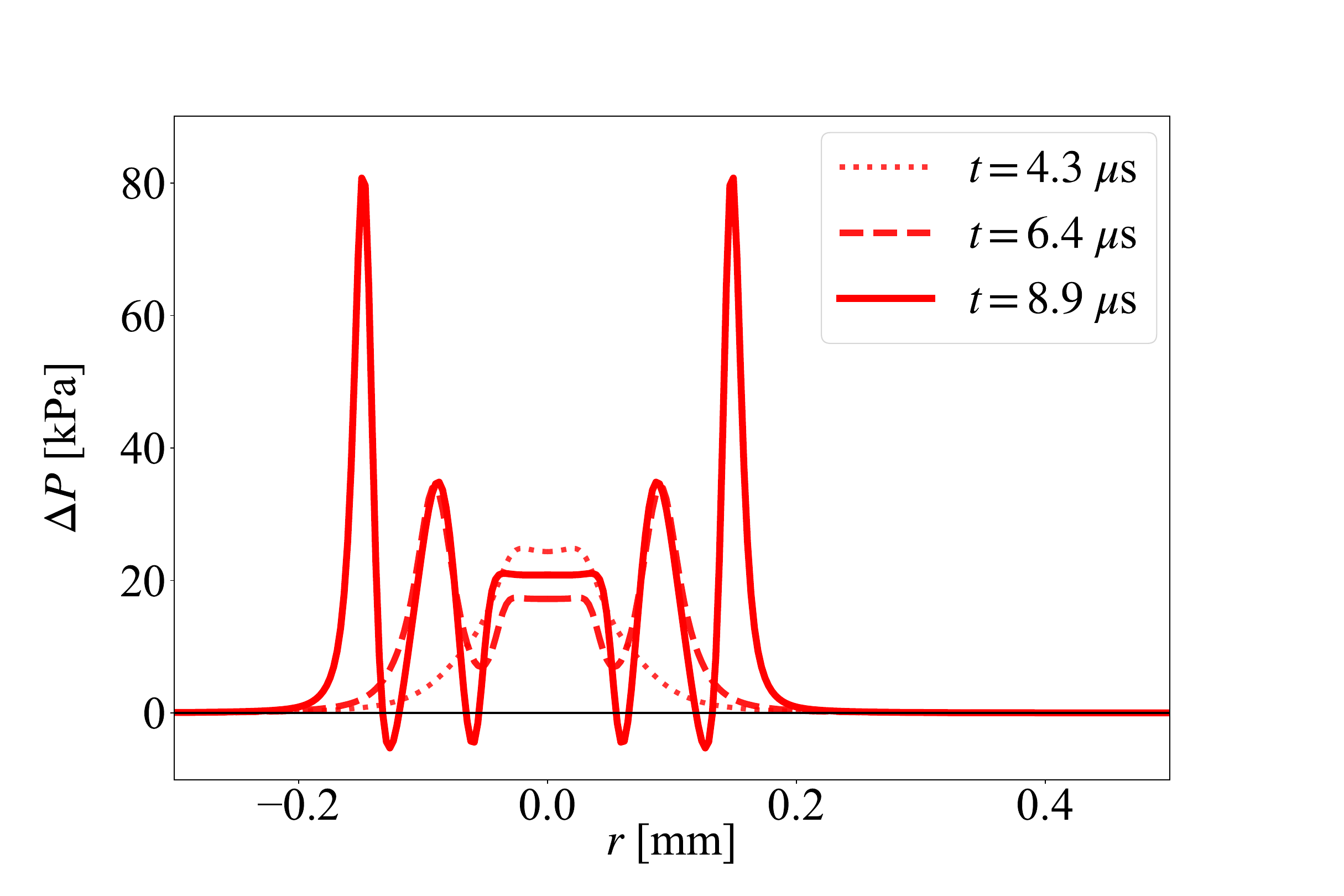}
\makebox{b)}
\caption{ a) Time evolution of the force $F(t)$ on the solid when a row of bubbles is at a distance $l = 5$ $\mu$m from the droplet's surface. The humps on the plot correspond to instances when the pressure build up is released due to a torus deforming. The snapshots from the BI simulation show the region of impact at $t = 4.3$, 6.4 and 8.9 $\mu$s where both the deformation of the droplet interface and that of the bubbles are clearly discernable. b) Pressure profile $\Delta P (r,t)$ at $t = 4.3$, 6.4 and 8.9 $\mu$s, i.e., the same times at which the snapshos in (a) were taken, and which approximately correspond to the location of the humps in the force curve. The troughs in $\Delta P$ coincide with the positions where local maxima in $h_d$ appear, and are close to a deforming bubble above it. The peaks correspond to the positions where there are local minima in the separation distance between droplet and solid.} \label{fig:2_fig11}
\end{figure}

In order to explain the stronger force attenuation and fly time, let us look deeper into the case $l = 5 \; \mathrm{\mu m}$. Figure \ref{fig:2_fig11} (a) isolates the force exerted on the solid and shows snapshots of the impact zone at three specific moments in time, at $t = 4.3, 6.4$ and $8.9 \; \mathrm{\mu s}$. These correspond to humps or bulges in $F(t)$, indicating sudden changes in the time rate of change of the force. Qualitatively what happens is that, as the droplet approaches the solid and the pressure in the film rises, a first peak appears in $\Delta P$, which flattens once the droplet's surface starts to deform. Then, two peaks arise at the spot where there is minimum separation between the droplet and the solid. However, when more than one bubble is close to the surface, the dynamics do not stop there. The \textit{area of influence}, i.e., the bottom region where the pressure is high enough to deform the droplet in a visible way, continues to spread.

As the area of influence expands, it reaches the radial distance where directly above lies a second bubble ($\mathrm{T_1}$). On this ring (recalling the axisymmetry of the simulation), the droplet presents less resistance to deformation and a local dimple is created. Once this happens, some of the pressure built up in the peaks is released due to the formation of the second dimple, decreasing the pressure locally. This \textit{slows down} the rate at which the load on the solid increases. A local minimum in $h_d$ (corresponding to a maximum in $\Delta P$) appears once the area of influence reaches a radial distance where just above there is no bubble and neighboring bubbles have been pressurized, thus creating a second pair of peaks in the pressure profile. The process goes on as the impact area of influence continues to spread and reaches the second ($\mathrm{T_2}$) and third ($\mathrm{T_3}$) torus. Each time the torus deforms, it releases pressure and the fly time increases. 

Figure \ref{fig:2_fig11} (b) presents the pressure profile $\Delta P (r,t)$ at the instants in time $t$ where the humps in $F(t)$ arise, corresponding to the simulation snapshots in Fig.~\ref{fig:2_fig11} (a). The dotted line ($t = 4.3 \; \mathrm{\mu s}$) shows the moment when the first pressure peak flattens and a first pair of horns starts to form. The dashed line ($t = 6.4 \; \mathrm{\mu s}$) corresponds to the time at which the first pair of horns starts to flatten and a second pair begins to arise. At $t = 8.9 \; \mathrm{\mu s}$ (solid line) the process repeats once again. 

\begin{figure}
\centering
\includegraphics[trim={5 15 45 15},clip,width=0.99\columnwidth]{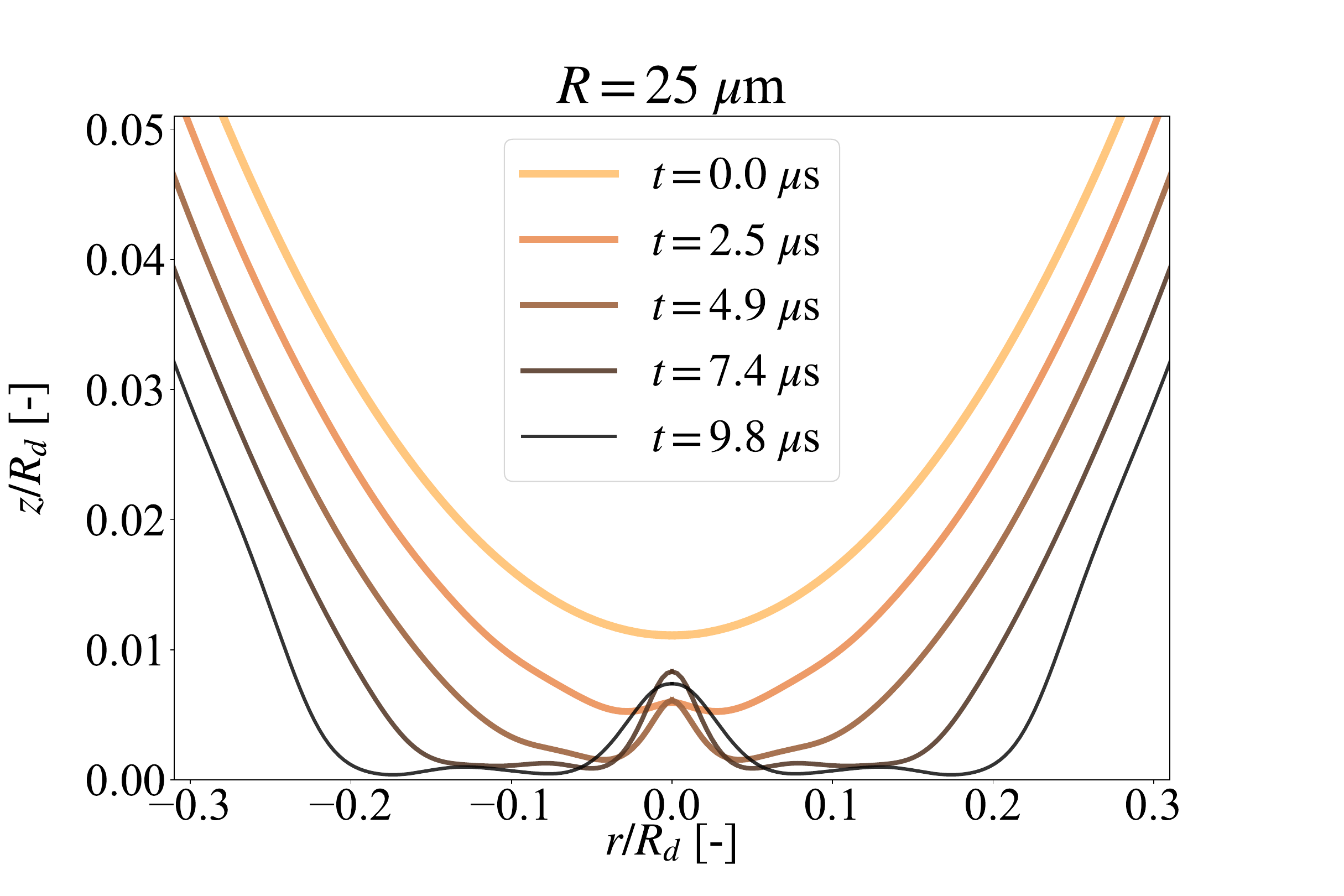}
\caption{Vertical zoom in into the droplet surface when there is a full row of bubbles separated at a distance $l = 5 \; \mathrm{\mu m}$ to the droplet surface. Bubbles are not shown for clarity. The lighter colors correspond to initial stages of the impact and the darkest color to the moment just before touchdown. It is observed that multiple dimples appear which spread as time passes by. 
}\label{fig:2_fig11_1}
\end{figure}

Finally, in Fig. \ref{fig:2_fig11_1} we vertically zoom in on the shape of the bottom surface of the droplet, at different times $t$. Note that, for clarity, the bubbles have not been depicted in this plot. Here, one clearly observes the formation and growth of the main dimple at the center (visible at $t = 4.9 \; \mathrm{\mu s}$), the spreading of the area of influence and the formation of the subsequent local minima (visible at $t = 7.4 \; \mathrm{\mu s}$ and $t = 9.8 \; \mathrm{\mu s}$), consistent with the evolution of events discussed above. In addition, in the curve corresponding to the latest time ($t = 9.8 \; \mathrm{\mu s}$), it is seen that subsequent minima in the radial direction become deeper, closer to the substrate, corresponding to a larger pressure maximum. As a result, eventually the contact between liquid and substrate will be established at the outermost minimum, which determines the size of the bubble that will be entrapped below the impacting droplet.

\subsubsection{Fixing $l$ and $R$, increasing number of bubble rows} 

The effect of one full row of bubbles (1S 3T) has been investigated in the last subsection. This final subsection is dedicated to what is depicted in Fig. \ref{fig:2_fig3} (f), where more than one, vertically stacked, full rows of bubbles are simulated. The first row is placed at a distance $l = 10 \; \mathrm{\mu m}$ from the droplet's surface, and again all bubbles have a radius of $R = 25 \; \mathrm{\mu m}$ and are separated from each other by 0.07 rad. The second row of bubbles is placed directly above the
\textcolor{black}{first row at a distance of $2R+l$ measured from the center of the spherical bubble of the first row to the center of the spherical bubble of the second row. The third row is placed above the second one at the same separation and following the same idea, i.e., the separation distance between the center of the spherical bubble of the second row and the center of the spherical bubble of the third row is $2R+l$.}

\begin{figure}[t]
\centering
\includegraphics[trim={25 15 25 15},clip,width=0.99\columnwidth]{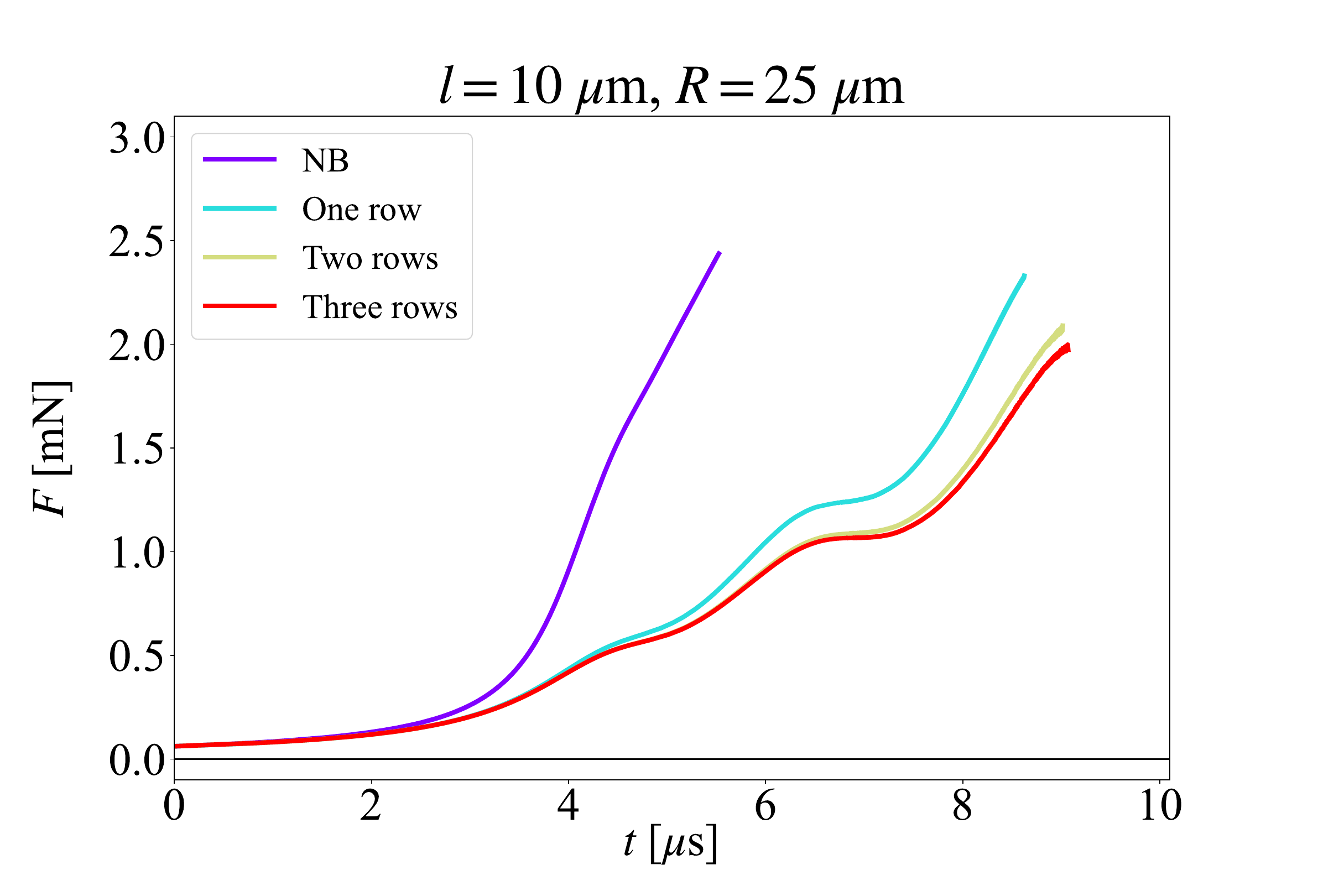}
\makebox{a)}
\includegraphics[trim={15 15 35 15},clip,width=0.99\columnwidth]{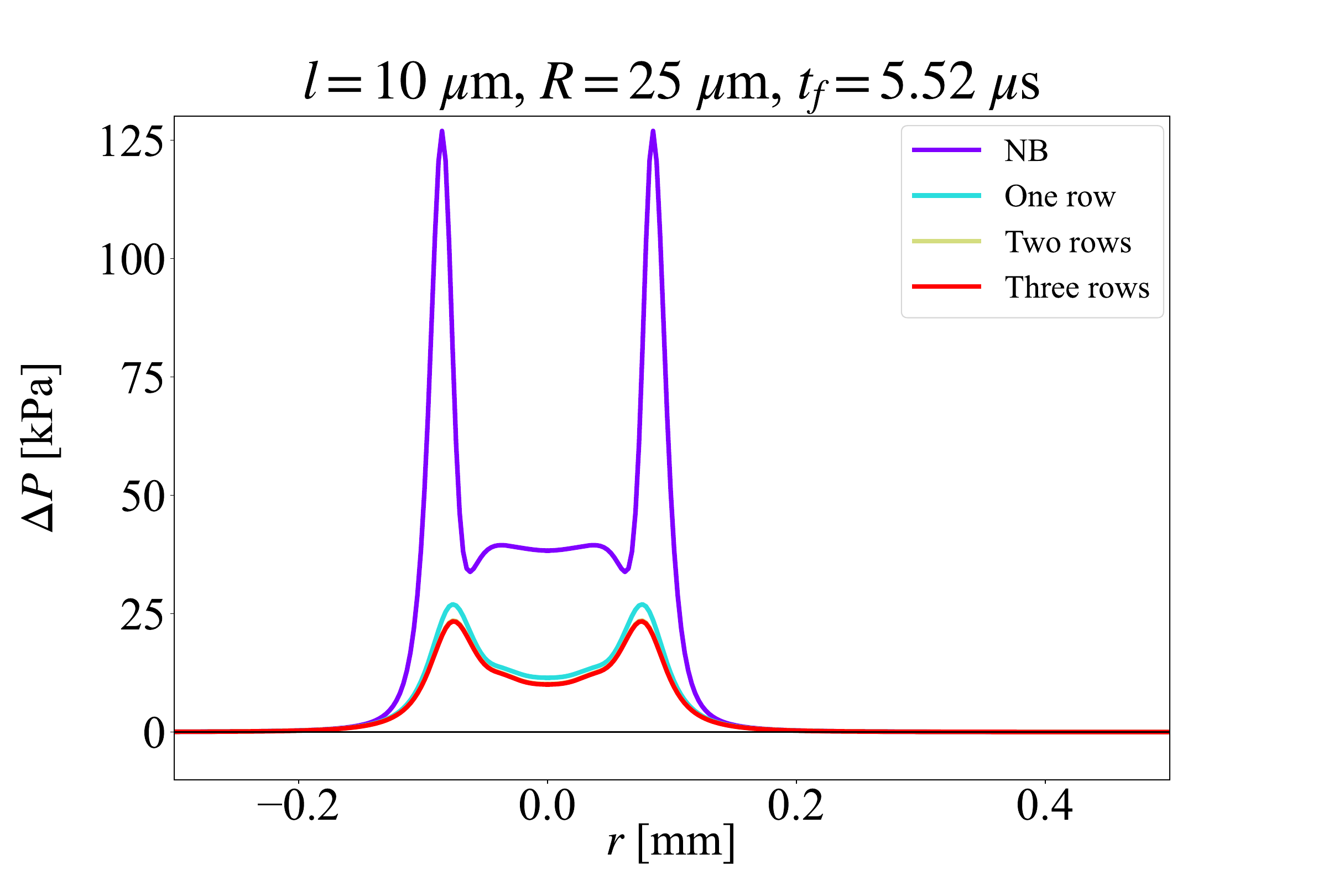}
\makebox{b)}
\caption{ a) Time evolution of the force $F(t)$ exerted on the solid substrate for different numbers of full rows stacked on top of each other. Although there is an appreciable difference between one and two rows, $F(t)$ is practically identical for the cases with two and three rows since especially the third row is far from the impact zone. b) Pressure profiles $\Delta P (r,t_f)$ at $t_f = 5.52$ $\mu$s. Again, the profile for two and three rows is virtually identical.}\label{fig:2_fig12}
\end{figure}

Figure \ref{fig:2_fig12} (a) and (b) once again show the large attenuation effect that one row of bubbles can produce as compared to the NB case which can be observed by looking either at the force or the pressure profile. In contrast, it is remarkable that the difference between one and two rows is minor, and between two and three rows is minimal. The similarity between the cases containing two and three rows again originates from the region of influence: While the first and second row take part in the deformation, row three is simply too far away from the impact zone to have a significant effect.

\section{Conclusions}\label{sec:2_4}

The purpose of this study is to numerically study the effect of bubbles present inside a droplet on the pressures and forces that occur during the first stages of its impact onto a solid substrate. To this end, the BI method was applied to simulate the initial stages of droplet impact onto a solid and to investigate the short-time force response of the substrate just before touchdown. Lubrication theory was used to model the layer of air between the droplet and solid. An expression for the pressure in the air film was derived and employed to compute the load on the solid surface. In addition, bubbles were implemented inside the droplet simulation and modeled using a polytropic equation of state, which allows for compression and expansion of the gas inside of them. It was found that gas bubbles inside the impacting droplet mitigate the load exerted onto the solid. Subsequently, a variety of systems were simulated to explore the influence of the bubble's size, position, number and shape (spherical and toroidal) on the pressure and thus, force on the solid.

It was found that bubbles that are close to the surface add compressibility and deformability to an otherwise incompressible droplet. This cushions the impact considerably more than a bubble further away from the impact zone. The bubble size is also of importance in the cushioning effect, since larger bubbles are able to deform and compress more before collapsing. It was found that inertial effects (reduced liquid mass due to bubbles) are negligible as long as the bubble's volume is not comparable to that of the droplet.

Stacking bubbles vertically, either in the form of spherical bubbles or full rows, does not have a big effect on the load once these are far from the impact zone, as long as inertial effects can be neglected (reduction of liquid mass due to bubbles inside the drop).

It is also found that local minima and maxima appear in the air layer width when toroidal bubbles are placed close to the surface. This creates multiple rings of peaks in the pressure profile. The release in pressure due to bubble deformation reduces the rate at which the force exerted on the solid increases. Several tori inside the droplet deform the surface in a wavy shape rather than the sharper dimple-like way in which the surface is deformed for the case without or with a single bubble. This significantly increases the fly time of the droplet. 

\begin{acknowledgements}
This publication is part of the Vici project IMBOL (Project No. 17070) which is partly financed by the Dutch Research Council (NWO). 
\end{acknowledgements}

\nocite{*}
\bibliography{stfrodiwgbbib}

\appendix

\section{Lubrication approximation}\label{sec:2_5}

In order to obtain an equation for the pressure inside the air layer we start from the radial component of the \textcolor{black}{incompressible} Navier-Stokes equations in cylindrical coordinates. Using lubrication approximation \cite{oron1997long} we can write
\begin{equation}
    \diffp{p}{r} = \mu_g 
    \diffp[2]{u_r}{z} \,\,.
\end{equation}
In this approximation, the vertical component reduces to $\partial p/\partial z = 0$, such that the pressure gradient does not depend on $z$ allowing us to directly integrate the above equation for $u_r$ with the boundary condition at the solid boundary $u_r (z=0) = 0$ and at the surface of the drop $u_r (z=h_d) = u_d$, where $u_d$ constitutes the radial velocity component at the droplet surface and $h_d(r,t)$ is the vertical distance of the droplet surface to the substrate. This leads to
\begin{equation}
    u_r = u_d \left( \dfrac{z}{h_d} \right) +
    \dfrac{1}{2 \mu_g} \diffp{p}{r} (z^{2} - h_d z )\,\,. 
    \label{eq:2_6}
\end{equation}
On the other hand, continuity dictates that
\begin{equation}
    \dfrac{1}{r}\diffp{(ru_r)}{r} + \diffp{u_z}{z} = 0 ,
\end{equation}
which one may integrate over the axial coordinate $z$
\begin{equation}
    \int^{h_d}_{0} \dfrac{1}{r}\diffp{(ru_r)}{r} \mathrm{d}z +
    u_z \bigg{|}_{z=h_d} = 0\,\,,
\end{equation}
where we have used that $u_z$ vanishes at the substrate ($z=0$). Combined with the kinematic boundary condition
\begin{equation}
    \textcolor{black}{u_z \bigg{|}_{z=h_d} = \diffp{h_d}{t} + \diffp{h_d}{r} u_r \bigg{|}_{z=h_d}} 
    \,\,, 
\end{equation}
we arrive at
\begin{equation}
  \dfrac{1}{r}\dfrac{\partial}{\partial r}\left[r \int_{0}^{h_d}  u_r \mathrm{d} z\right]
    + \dot{h}_d = 0 \,\,, \label{eq:2_12}
\end{equation}
where $\dot{h}_d$ is short for 
\textcolor{black}{$\partial h_d/\partial t$} at the interface. We can substitute $u_r$ from Eq. \eqref{eq:2_6} into \eqref{eq:2_12}, perform the integration and multiply by $r$ to find
\begin{equation}
\diffp{}{r}\!\left[ r\left( \tfrac{1}{2}u_d h_d - \frac{h_d^3}{12 \mu_g} \diffp{p}{r} \right)\right] + r \dot{h}_d = 0\,\,.
\end{equation}

Integrating the above expression with respect to $r$ and solving for $\partial p/\partial r$ we arrive at
\begin{equation}
    \diffp{p}{r} = 
    \frac{ 12 \mu_g }{ r h_d^{3} } \int_{0}^{r} \tilde{r} \dot{h}_d \mathrm{d} \tilde{r} + \dfrac{ 6 \mu_g u_d }{ h_d^2 }\,\,,\label{eq:A8}
\end{equation}
which can be numerically integrated every time step to obtain the pressure beneath the droplet inside the air layer.

\begin{figure}[t]
\centering
\includegraphics[trim={3 15 25 15},clip,width=0.99\columnwidth]{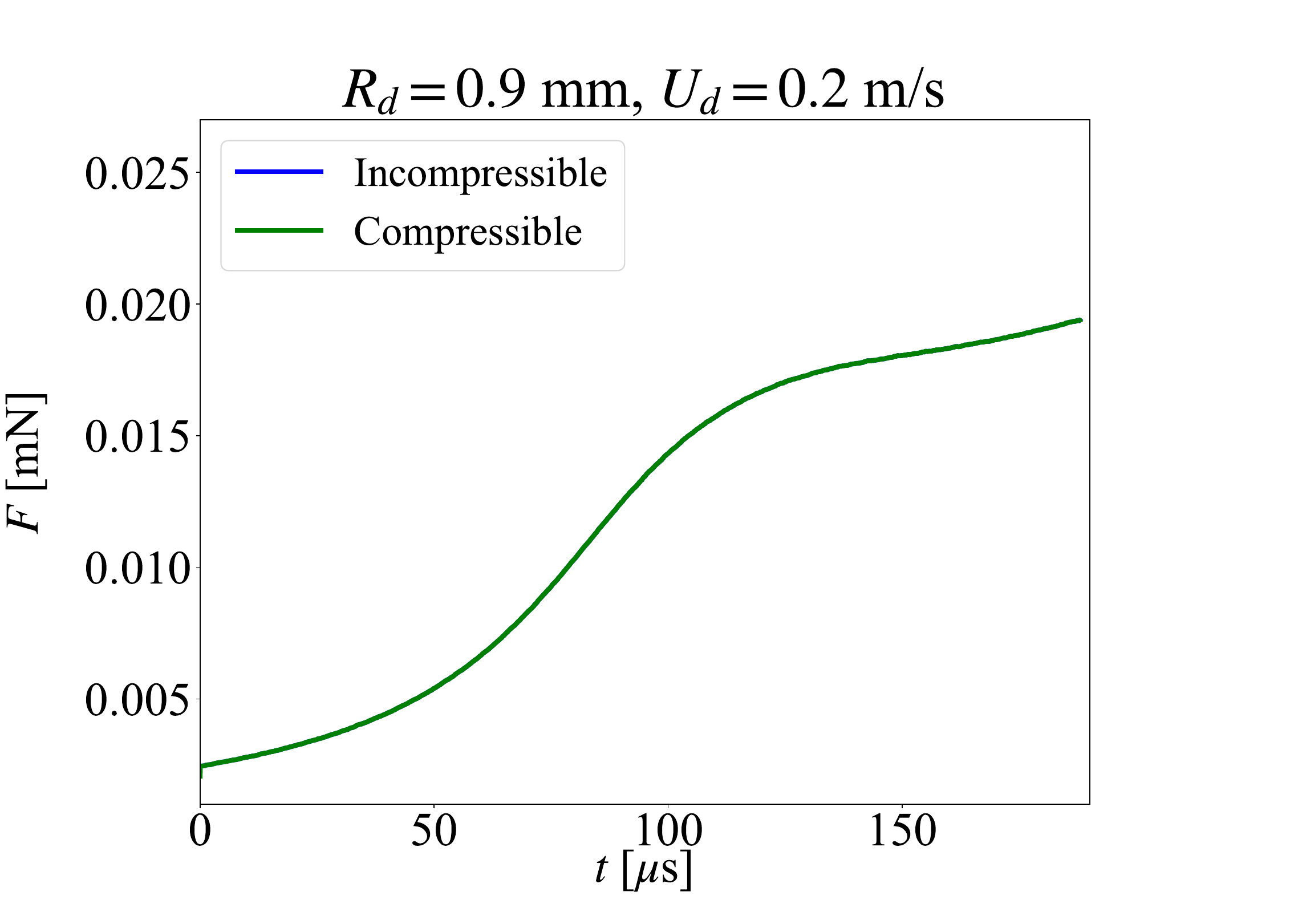}
\makebox{a)}
\includegraphics[trim={3 15 25 15},clip,width=0.99\columnwidth]{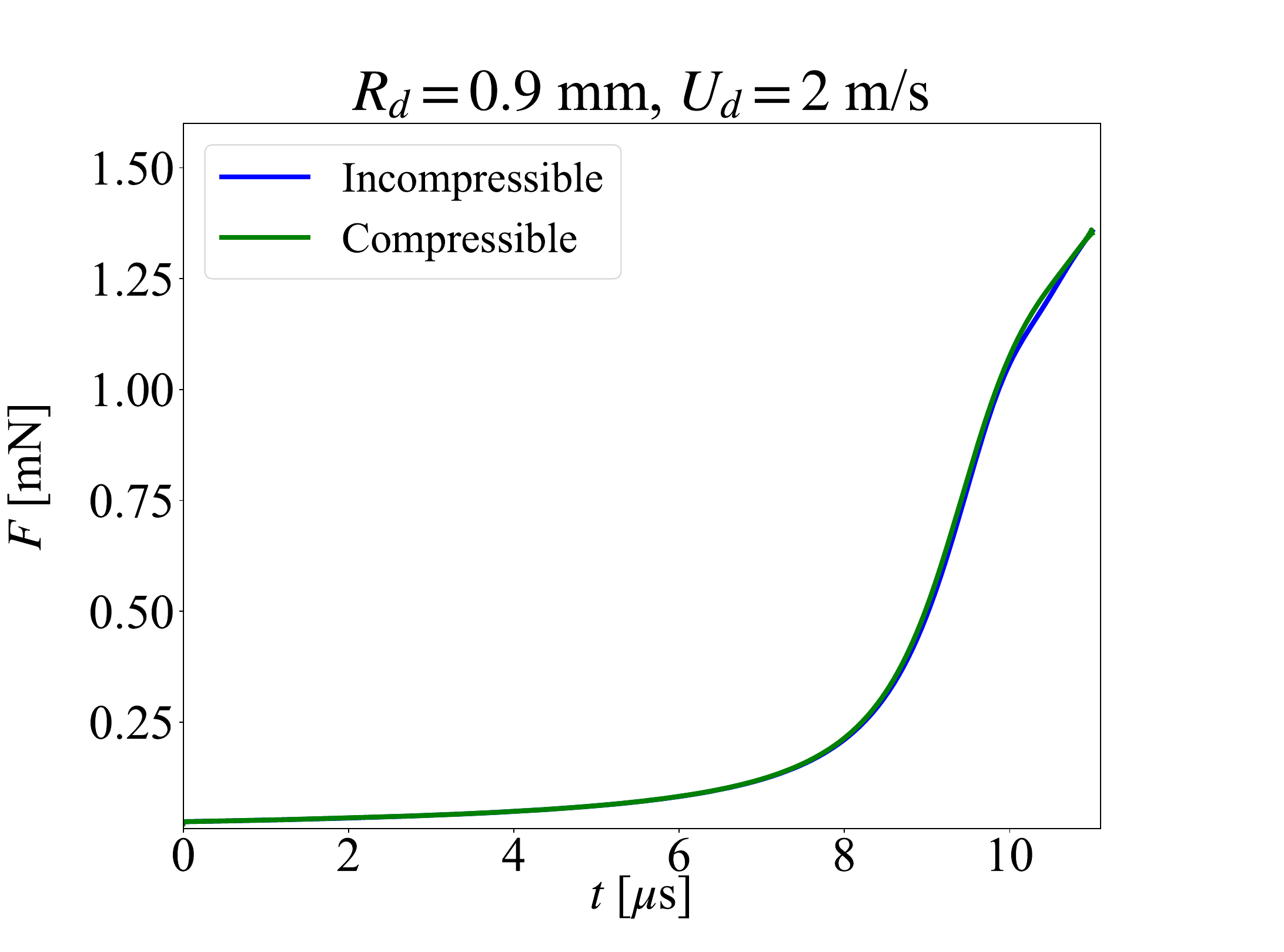}
\makebox{b)}
\includegraphics[trim={3 15 25 15},clip,width=0.99\columnwidth]{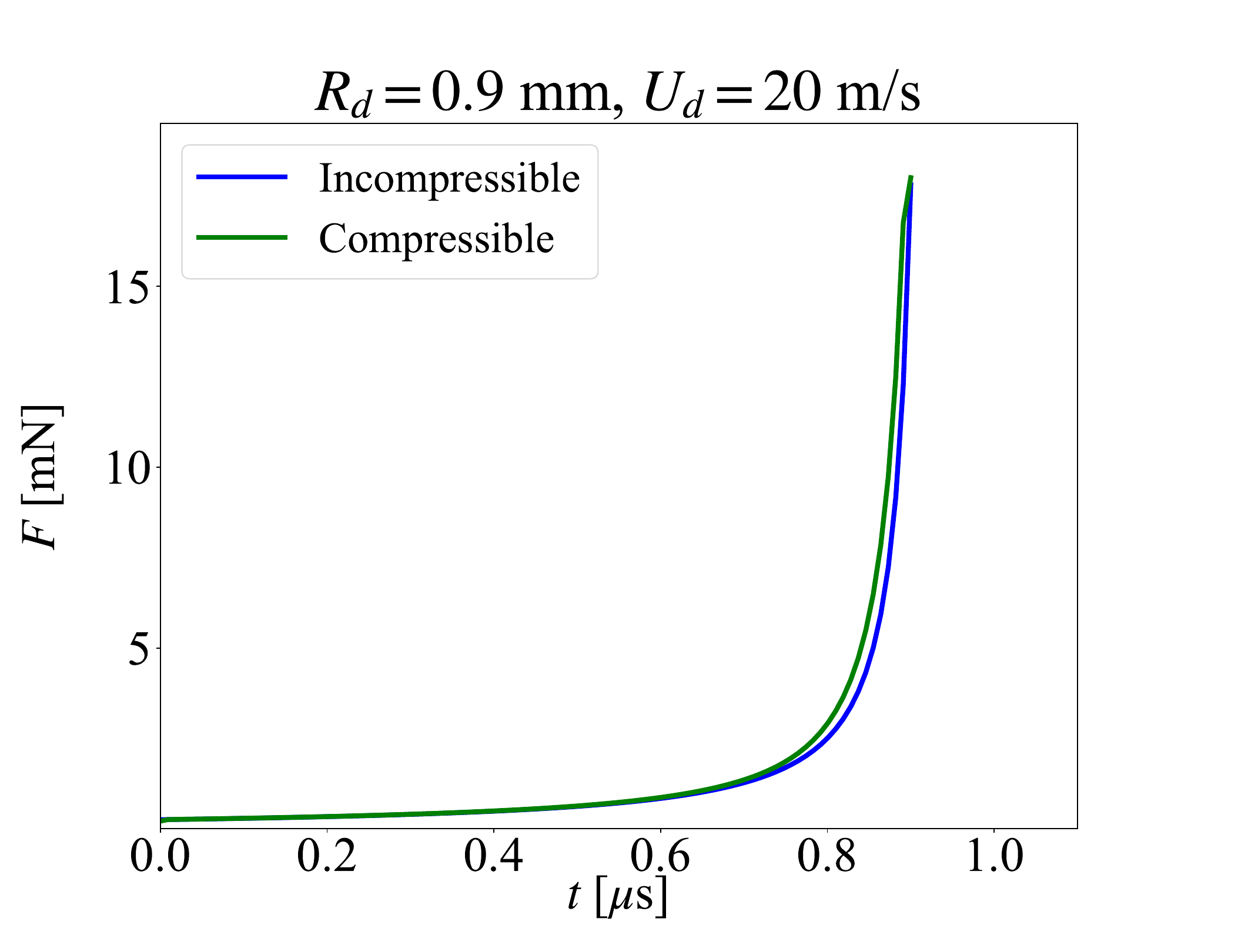}
\makebox{c)}
\caption{\textcolor{black}{Time evolution of the force $F(t)$ that the solid substrate experiences as a droplet of $R_d = 0.9$ mm (without any bubbles inside) approaches the substrate at a) $U_d = 0.2$ m/s, b) $U_d = 2$ m/s and c) $U_d = 20$ m/s. The cases where the lubrication layer is modelled as incompressible are plotted in blue, and the cases where the lubrication layer is modelled as compressible are plotted in green.}} \label{fig:2_fig15}
\end{figure} 

\textcolor{black}{Finally, an analogous compressible formulation of the lubrication layer is equally possible, in which case the problem needs to be supplemented with an equation of state for the gas phase, e.g., expressing adiabatic compression. The numerical problem attached with such an approach, however, is connected to the fact that an additional time derivative of the density appears in the integral in \eqref{eq:A8}, which is depending on the time derivative of the pressure which in turn needs to be obtained from comparing multiple timesteps of the boundary integral part of the simulation. This introduces additional numerical errors that need to be dealt with and substantially increase computation time.} 
    
\textcolor{black}{Nevertheless, we have performed such simulations for selected cases and have found differences between the compressible and incompressible approach to be small for the parameter settings used in this experiment, such that we decided to stick with the incompressible approach.} 

\textcolor{black}{ In Fig. \ref{fig:2_fig15}, we have plotted the time evolution of the force $F(t)$ on the solid when a droplet of $R_d = 0.9$ mm approaches the substrate at (a) $U_d = 0.2$ m/s, (b) $U_d = 2$ m/s and (c) $U_d = 20$ m/s respectively. The droplet does not contain any bubbles for the aforementioned cases, however we have also run simulations where a bubble is present inside the liquid and found that the difference between incompressible and compressible lubrication layer is minimal for the velocity range explored in this manuscript.}
  
\end{document}